\documentclass[epj]{svjour} 
\usepackage{graphicx}
\usepackage{amsmath,amsfonts,amssymb,array}

\def\p{\partial}
\def\cc{\centering \par}

\begin{document}
\title{Non-spherical shapes of capsules within a fourth-order\\ curvature model}

\author{O.~V. Manyuhina\inst{1} \and J.~J. Hetzel\inst{2} \and M.~I. Katsnelson\inst{2} \and A. Fasolino\inst{2}} 

\institute{\inst{1}{Laboratoire de Physique Statistique, Ecole Normale Superieure, UPMC  
Univ Paris 06, Universit\'e Paris Diderot, CNRS,\\ 24~rue Lhomond, 75005  
Paris,  France}\\ \inst{2}{Radboud University Nijmegen, Institute for Molecules and Materials, Heyendaalseweg 135, 6525 AJ Nijmegen, The~Netherlands}}


\abstract{We minimize a discrete version of the fourth-order curvature based Landau free energy by extending Brakke's Surface Evolver.  This model predicts spherical as well as non-spherical shapes with dimples, bumps and  ridges to be the energy minimizers. Our results suggest that the buckling and faceting transitions, usually associated with crystalline matter, can also be an intrinsic property of non-crystalline membranes.}%

\maketitle

\section{Introduction}
The study of organized structures, like self-assembled membranes and vesicles, is an important part of soft matter physics that is relevant for chemistry and biology and for practical applications~\cite{Jonesbook,grzybowski:2002}.
These systems can usually be described as two-dimensional surfaces, since their thickness is much smaller than the typical size. Thus, at the phenomenological level, the search for equilibrium shapes of such micro- and nano-structures is a problem of differential geometry of surfaces. This view is certainly appropriate for liquid membranes where the free energy is considered as a functional of the local curvatures and does not depend on the in plane deformation tensor like for crystalline membranes~\cite{nelsonbook}.

A simple theory for elastic shells was proposed by Sophie Germain around 1810~\cite{hsu:1992,germainbook} in which the energy is given by the form
\begin{equation}\label{eq:FSG}
E(S)=\iint dS\,\{\sigma + A H^2 + B K\},
\end{equation}
where $H$ is the mean curvature, $K$ is the Gaussian curvature and the integral is over the surface $S$. In the case of soap  films, the first term, proportional to the surface tension $\sigma$, is the dominant contribution to the free energy. For fluid membranes, instead,  $\sigma\simeq 0$ because molecules can easily flow and adjust the total area of the membrane to the one corresponding to the best packing~\cite{nelsonbook}. The second term in the integrand in Eq.~(\ref{eq:FSG}) describes out-of-plane bending of elastic shells. According to the Gauss--Bonnet theorem~\cite{docarmobook} for compact surfaces, the last term $\iint dS\,K=2\pi\chi(S)$ is a topological invariant, with $\chi(S)$ called the Euler--Poincar\'e characteristic of the surface. For all surfaces topologically equivalent to spheres $\chi=2$ and $\iint dS\,K=4\pi$.

The squared mean curvature integral $W=\iint dS\,H^2$ is known in mathematics as Willmore functional~\cite{willmore:1965}, in the theory of membranes as Helfrich free energy~\cite{helfrich:1973}, and in string theory as Polyakov action~\cite{nelsonbook}.  The sphere turns out to give the absolute minimum of this functional, $W=4\pi$, among all compact surfaces~\cite{willmore:1965}. This naturally explains why spherical shapes are common in the world of fluid membranes. More complicated equilibrium shapes than the spherical one, e.g. prolates, discocytes and stomatocytes~\cite{lipowsky:1991,seifert:1997},  can be obtained within the Willmore functional by adding the condition of constant volume.  Complicated equilibrium shapes are also observed in systems, like red blood cells or viral capsids,  where other degrees of freedom rather than bending become important~\cite{nelsonbook} so that the functional $W$ and thus Eq.~(\ref{eq:FSG}) is not sufficient. These situations are usually described in terms of crystalline membranes where these additional degrees of freedom are described by the in-plane deformation tensor. 

In Ref.~\cite{fasolino:2006},  for symmetric bolaamphiphilic fluid membranes, it was found that the interactions of the hydrophilic tails of the bolaamphiphiles with molecules of the solvent as well as entropic terms may lead to a negative coefficient $A$ in Eq.~(\ref{eq:FSG}). Then, symmetry allowed higher order terms should be added to stabilize the free energy that otherwise would not be bounded from below
leading to the free energy functional 
\begin{equation}\label{eq:F4}
{\cal F}_4=\iint dS\,\{-A H^2+B K+c_1H^4+c_2H^2K+c_3K^2\},
\end{equation}
where the minus sign is written explicitly so that from now on $A$ is positive.  
The interplay between these higher order terms and the Willmore functional with negative sign leads to spontaneous bending.  This model was successfully applied to describe the experimental data on deformations of spherical bolaamphiphilic vesicles in high magnetic fields~\cite{prl:2007}. Here we study the equilibrium shapes of liquid membranes with spontaneous bending based on minimization of the functional Eq.~(\ref{eq:F4}).

Without referring to any specific system,  we  minimize numerically ${\cal F}_4$ under the constraint of constant surface ($S=4\pi R^2$) and preserving the topology of the sphere ($\chi=2$). To this purpose we have supplemented the open-source software ``Surface Evolver''~\cite{brakke:1992} with new  subroutines for the calculation of fourth order terms. Increasing $A$, for a given set of $c_i$, we find a transition from a spherical shape to more complicated, dimpled, shapes when $AR^2/c_1\simeq 2$. This continuous transition is followed by a discontinuous one towards shapes of icosahedral symmetry with ridges and facets, when $AR^2/c_1>8$. The presence of the coupling term between the mean curvature and the Gaussian curvature ($c_2\neq0$) results in intermediate shapes with bumps. We characterize these new equilibrium shapes in terms of order parameters (rotational invariants), which were introduced to describe virus capsids~\cite{lidmar:2003} and bond-orientational order in liquids and glasses~\cite{steinhardt:1983}. Moreover, by comparing rotational invariants, we draw an analogy between the buckling transition found in virus capsids~\cite{lidmar:2003} and the transition from spherical shapes towards shapes with dimples and ridges explored in this paper.

\section{Phenomenological models}

The spontaneous-curvature model introduced by Helfrich in 1973~\cite{helfrich:1973} accounts for  a possible asymmetry of membranes, such as a difference in the number of molecules in each layer of bilayer vesicles. He suggested the following free energy of fluid membranes 
\begin{equation}\label{eq:FH}
{\cal F}_{\rm H}=\iint dS\, \{2\kappa(H-H_0)^2+\bar\kappa K\},
\end{equation}
where $H_0$ is called the spontaneous curvature,  $\kappa$ is known as bending rigidity and $\bar \kappa$ is the Gaussian rigidity that affects only transitions implying a change of topology. The Helfrich model was successful in describing different phenomena, like the budding transition, discocyte--stomatocyte transition, et cetera~\cite{lipowsky:1991,seifert:1997}. However, for some systems it turns out to be too simple and thus insufficient to explain new experimental data. In fact the Helfrich approach accounts only for the basic degrees of freedom, such as bending, neglecting the possibility of a tilt of the molecules within a layer~\cite{may:2000}, stretching/compression of layers~\cite{stewart:2007} and interaction with the environment~\cite{fasolino:2006}. The latter situation may result in a free energy with spontaneous bending (negative coefficient in front of $H^2$) given by Eq.~(\ref{eq:F4}). In general, one can imagine other mechanisms favouring spontaneous bending, like geometric constraints of packing, complex van der Waals or electrostatic intermolecular interactions.

In this paper we consider only symmetric fluid membranes, where the Helfrich free energy ${\cal F}_{\rm H}$ with $H_0\equiv0$ coincides with the Willmore functional  $W=\iint dS\,H^2$. It is worth noting that ${\cal F}_{\rm H}$  does not depend on the size of the structure. The presence of higher order terms in ${\cal F}_4$ (Eq.~(\ref{eq:F4})), on the contrary,  result in  a characteristic length scale, $L\propto \sqrt{c_i/A}$. To guarantee that ${\cal F}_4$ has a minimum for real values of $H$ and $K$ we require the form $\Phi(H^2,K)=c_1H^4+c_2H^2K +c_3K^2$ to be positive definite for $H^2\geqslant K$. Thus a minimum exists if 
\begin{equation}\label{eq:con1}
c_1>0,\quad c_3>0,\quad 4c_1c_3-c_2^2>0.
\end{equation}
To relate the coefficients of Eq.~(\ref{eq:F4}) to the bending rigidity $\kappa$, entering the Helfrich model Eq.~(\ref{eq:FH}), we calculate $\p^2{\cal F}_4/\p H^2$ at $H=H_0$ which provides a minimum of the functional. The result is
\begin{equation}\label{eq:kappa}
\kappa=\frac{4A\,c_1c_3}{4c_1c_3-c_2^2}>0.
\end{equation}
In the case $c_2=0$, $\forall c_1,c_3>0$ this expression simplifies to $\kappa=A$. In principle the parameters $c_i$ can be determined by comparing equilibrium shapes  with experimental values for the deformations due to high magnetic fields as it has been done for sexithiophene vesicles~\cite{prl:2007}.  However, the coefficients $c_i$ are not separately accessible. Here, we will consider them as formal parameters, satisfying Eq.~(\ref{eq:con1}) and study the possible equilibrium shapes and their transformations, without referring to specific systems.

\begin{figure*}[t]
\centering
\raisebox{32mm}{(a)}\kern-10pt\includegraphics[height=37mm]{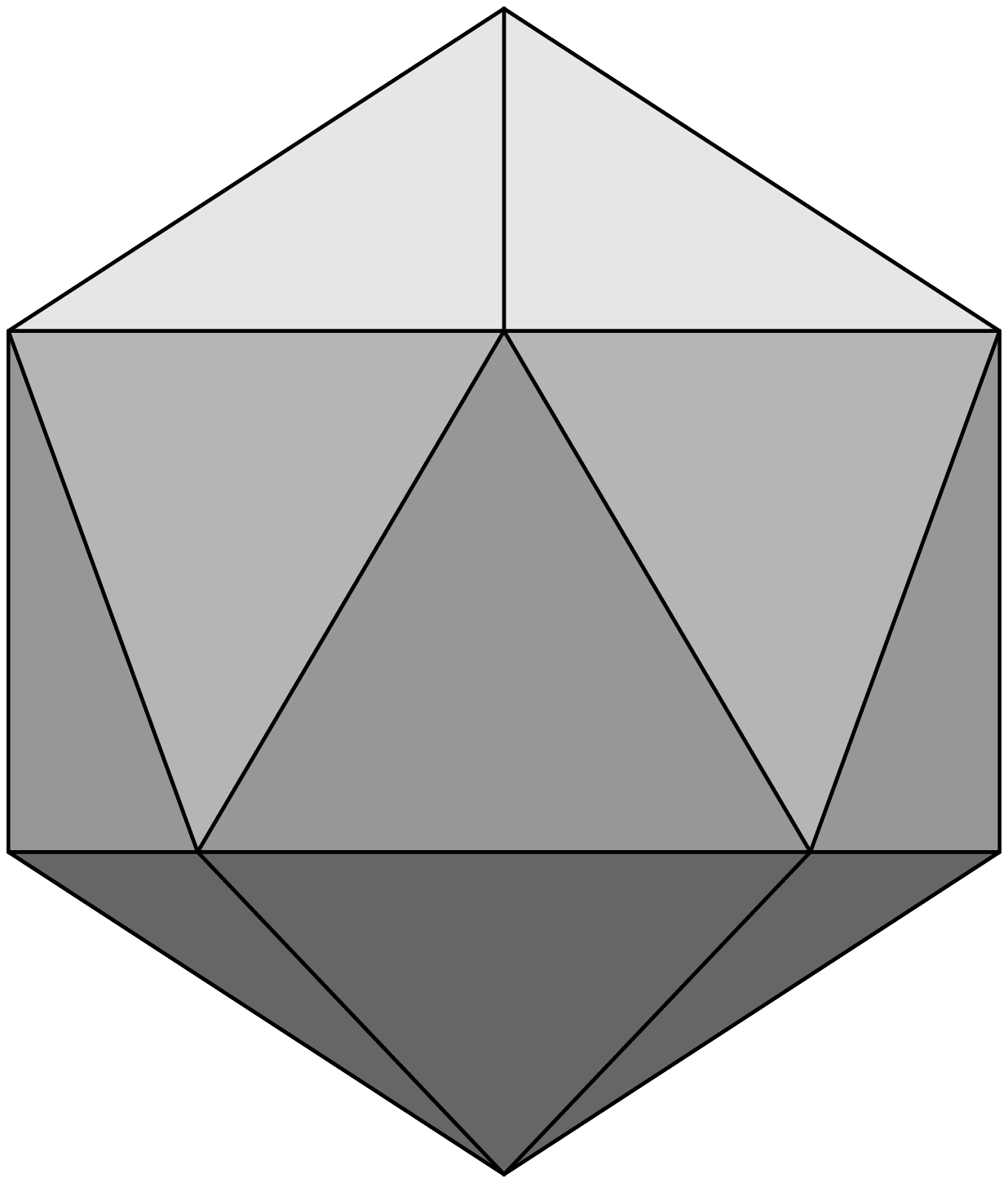}
\hfil\raisebox{16mm}{$\Longrightarrow$}\hfil
\raisebox{32mm}{(b)}\kern-10pt\includegraphics[height=37mm]{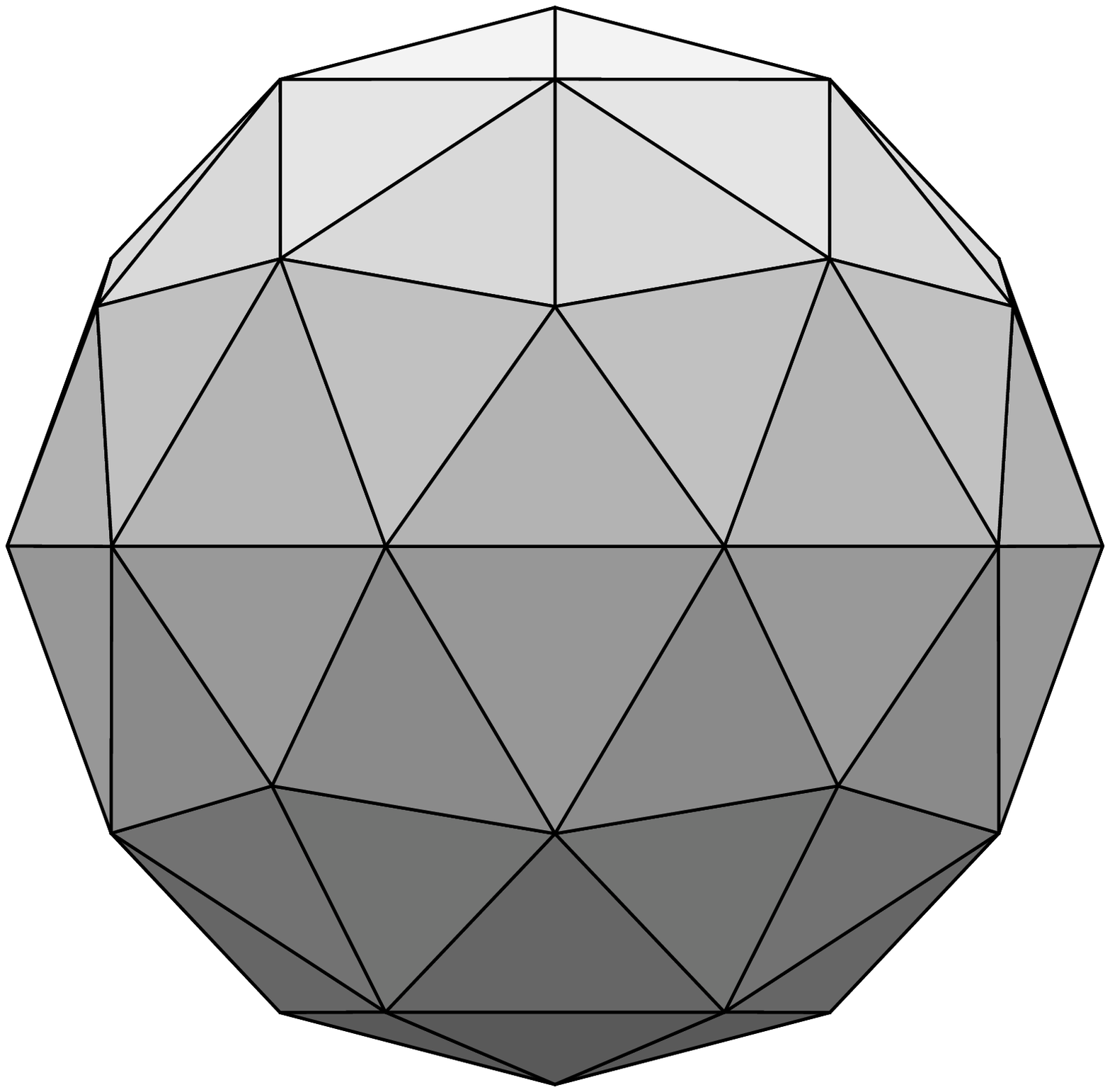}
\hfil\raisebox{16mm}{$\Longrightarrow$}\hfil
\raisebox{32mm}{(c)}\kern-10pt\includegraphics[height=37mm]{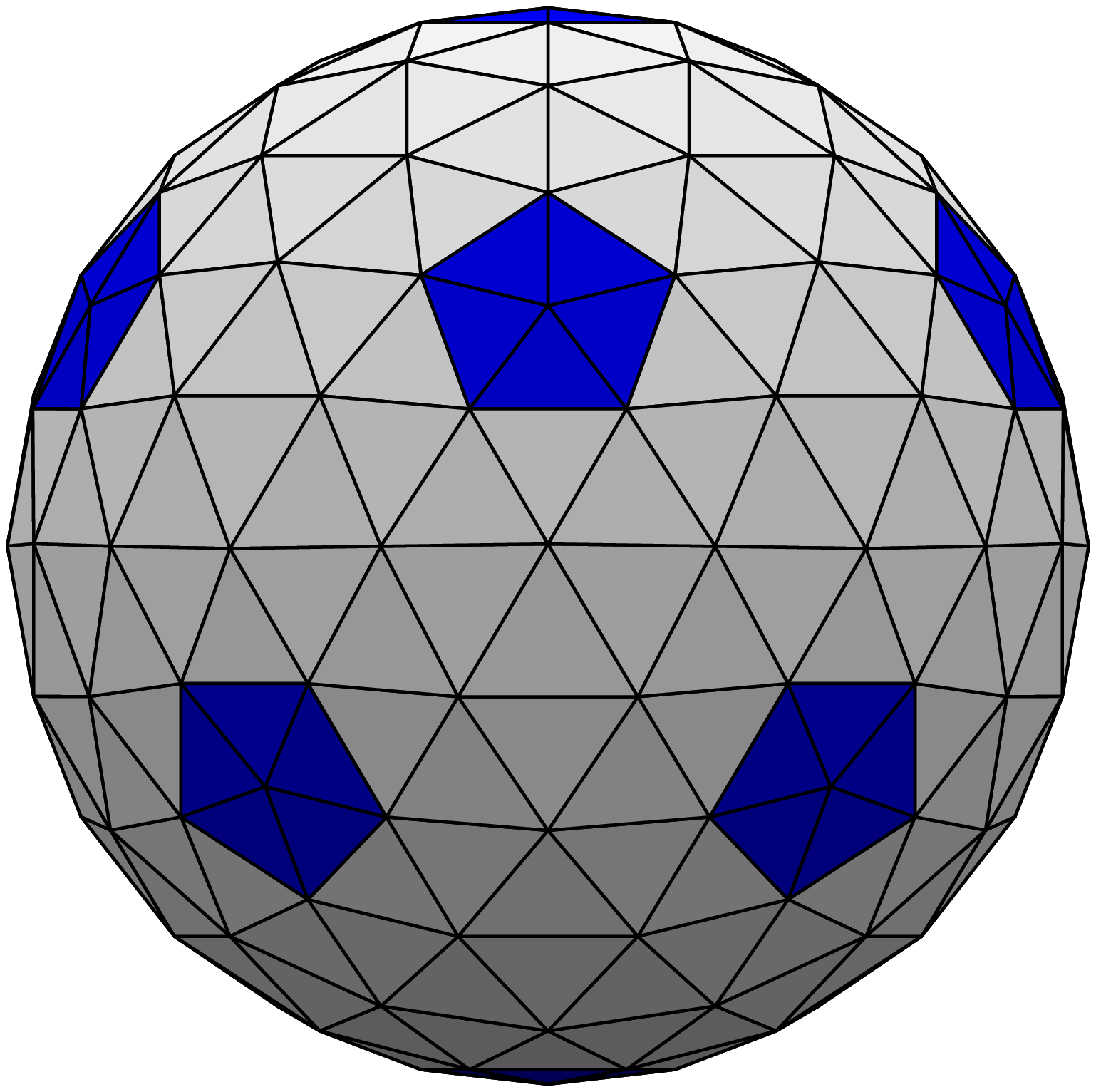}
\caption{An example of the evolution of icosahedron (a) towards a sphere (c) for the Willmore free energy. Every iteration, each triangle is subdivided into  four similar smaller triangles. The blue color on (c) illustrates the vertices  with a fivefold coordination.}
\label{fig:evolve}
\end{figure*}

To find equilibrium shapes, one usually needs to solve the Euler--Lagrange equation. For the Willmore functional $W$ only six analytic solutions are known:  planes, cylinders, spheres, tori, cones, dupin cyclides~\cite{zhongcanbook}. Since we do not consider here topological transformations, the sphere gives the absolute minimum $W=4\pi$ for $\chi=2$. The equilibrium shape equation for the functional ${\cal F}_4$ can be written following  Ref.~\cite{zhong-can:2001}, which results in
\begin{multline}\label{eq2:shape}
2AH(H^2-K)+4(c_2-c_1)H^3K+2(c_3-c_2)HK^2+6c_1H^5+\\+\nabla^2(AH+2c_1H^3+c_2HK)+\bar\nabla^2(c_2H^2+2c_3K)=0.
\end{multline}
Here $\nabla^2$ stands for the Laplace--Beltrami operator, $\nabla^2=(1/\sqrt{|g|})\partial_i \big(g^{ij}\sqrt{|g|}\partial_i \big)$, $g_{ij}$ is the metric tensor of the surface, $g^{ij}=(g_{kl})^{-1}_{ij}$, $|g|=\det||g_{ij}||$,    $\bar{\nabla}^2=(1/\sqrt{|g|}) \p_i(KL^{ij}\break\sqrt{|g|}\p_j)$, $L_{ij}$ is a second fundamental form~\cite{docarmobook}. By substituting $H^2=K=1/R^2$, where $R$ is the radius of a sphere, into Eq.~(\ref{eq2:shape}), one finds a sphere as  solution if and only if $c_1+c_2+c_3=0$,  which contradicts the condition~(\ref{eq:con1}). Therefore a sphere is not an equilibrium shape of the free energy (\ref{eq:F4}). Finding analytical solutions for a highly nonlinear partial differential equation (Eq.~(\ref{eq2:shape})) is not likely. Alternatively, one can study equilibrium shapes by means of  numerical methods. Here we adapt the  ``Surface Evolver''  software~\cite{brakke:1992,susqu} to this purpose.


\section{Computational details}

In the interactive program ``Surface Evolver'', a surface is modeled by a set of triangles, which is finite for compact surfaces. Given a triangulation of the surface we evolve it towards the shape that minimizes the total free energy.  For the evolution ``Surface Evolver'' uses the steepest descent method, which means that at each iteration all vertices are moved along the gradients of the free energy. Then we can refine the surface by dividing each triangle into four, and repeat the procedure until the approximated surface becomes smooth. An example of the evolution, starting from the icosahedron, towards a sphere is shown in Fig.~\ref{fig:evolve}. Since the Euler characteristic $\chi$ does not depend on the triangulation of the surface, but only on the topology, it always holds $\chi = F-E+V=2$, which relates the total number of triangles (faces $F$), edges ($E$) and vertices ($V$). 

The first application of  ``Surface Evolver'' was to study the shape minimizers of the Willmore functional $W$, starting from polyhedra with different $\chi$~\cite{hsu:1992}. The Helfrich spontaneous curvature model (Eq.~(\ref{eq:FH})) with $H_0\neq0$ and volume constraint was analyzed with this program in~Ref.~\cite{zhong-can:1998}, resulting in the prediction of corniculate, tubelike and other nonaxisymmetric shapes of vesicles. Among other examples are the study of the rheology of foams, simulation of microgravity, phenomena of capillarity and wetting~\cite{susqu}.

Here we aim at studying the energy minimizing shapes of the free energy (\ref{eq:F4}), starting, for reasons that we explain later, from an icosahedron (see Fig.1a). Surface Evolver v.2.30 evaluates the energy terms, $\iint dS\,H^2$ and  $\iint dS\,K^2$. We have written two new subroutines to calculate the other two terms entering Eq.~(\ref{eq:F4}), namely $\iint dS\,H^4$ and $\iint dS\,H^2K$~\cite{routines}. For discrete surfaces at each vertex $\nu$, the mean curvature $H_\nu$ and the Gaussian curvature $K_\nu$  are defined as~\cite{meyer:2003,sullivan:2006}
\begin{equation}
H_\nu \equiv\frac{|\nabla S_\nu|}{2S_\nu},\qquad K_\nu \equiv \frac1 S_\nu (2\pi - \sum_i\phi_{i,\nu}),
\end{equation}
where $S_\nu$ is a Voronoi area around $\nu$, i.e. the area of the cell consisting of all points closer to $\nu$ than to any other vertex, $\nabla S_\nu$ is the gradient of $S_\nu$ at $\nu$, and $\sum_i \phi_{i,\nu}$ is the sum of all facet angles at the vertex $\nu$ of icosahedron. The definition of $H_\nu$ comes from the fact that the mean (extrinsic) curvature measures the variation of the area element, displaced along the normal, divided by the corresponding change of volume.  The definition of $K_\nu$ comes from the Gauss--Bonnet theorem for a Voronoi region. Then, we can approximate the integrals  by assigning their energy contributions to the vertices only. The integrals are calculated as a sum over all vertices $\nu$ of the curvature times the area around a vertex, which gives:
\begin{align}
\iint dS\,H^4 &=\sum_\nu \frac{S_\nu}3H_\nu^4 =\sum_\nu \frac{S_\nu}3 \bigg(\frac{3 \nabla S_\nu}{2 S_\nu}\bigg)^4,\label{eq:H4}\\
\iint dS\,H^2K &=\sum_\nu \frac{S_\nu}3H_\nu^2 K_\nu =\notag\\ 
&=\sum_\nu \frac{S_\nu}3 \bigg(\frac {3 \nabla S_\nu}{2 S_\nu}\bigg)^2\bigg(\frac 3{S_\nu}\big(2\pi-\sum_i\phi_{i,\nu}\big)\bigg).
\label{eq:H2K}
\end{align}
Note that, for the calculation of the energy contributions, the area associated with a vertex is taken to be $1/3$ of $S_\nu$, rather than  $S_\nu$. This approximation simplifies the calculations for ``Surface Evolver'' and works best for triangles close to equilateral. 
The choice of icosahedron as a starting shape guarantees that, at every iteration, we are close to equilateral triangulation so that the approximation of the integrals as in Eqs.~(\ref{eq:H4}) and (\ref{eq:H2K}) holds. Moreover,  among all regular polyhedra, the icosahedron has the ratio of the surface area over the enclosed volume closest to that of a sphere. It is in general convenient to start the minimization from a shape close to a sphere  because the steepest descent method implemented in ``Surface Evolver'' finds  only one stable local minimum, which is not necessarily a global one. Moreover, in nature many shapes, like viruses, are found to have an icosahedral symmetry and we will compare the predictions of our phenomenological model with the models developed for viral capsids, in particular the one discussed in~\cite{lidmar:2003}.

\section{Numerical results}
\begin{table*}[t]
\caption {\label{tab1}Summary of the equilibrium shapes for different values of the parameters in  ${\cal F}_4=\iint \!dS \{-AH^2+c_1H^4+c_2H^2K+c_3K^2\}$. In the rows from left to right, shapes evolve from spherical towards non-spherical ones with increase of the negative contribution $-A\iint dS\,H^2$. The number of the column, given in brackets, corresponds to the label of the points in  Fig.~\ref{fig:Q6Q10} and Fig.~\ref{fig:W6W10}. All the shapes have a constant area and triangulation with a number of vertices $V=642$.}
{\def\tabcolsep{3pt}
\centering
\begin{tabular}{m{15mm}m{24mm}m{24mm}m{24mm}m{24mm}m{24mm}m{24mm}m{0.1mm}}
\cc (point) &\cc (1) $A=0$ & \cc (2) $A=4$ & \cc (3) $A=8$  & \cc (4) $A=12$ & \cc (5) $A=16$ & \cc (6) $A=20$& \\[1ex]
\cc $c_2=0$\\$c_3=0$ & \includegraphics[scale=0.16]{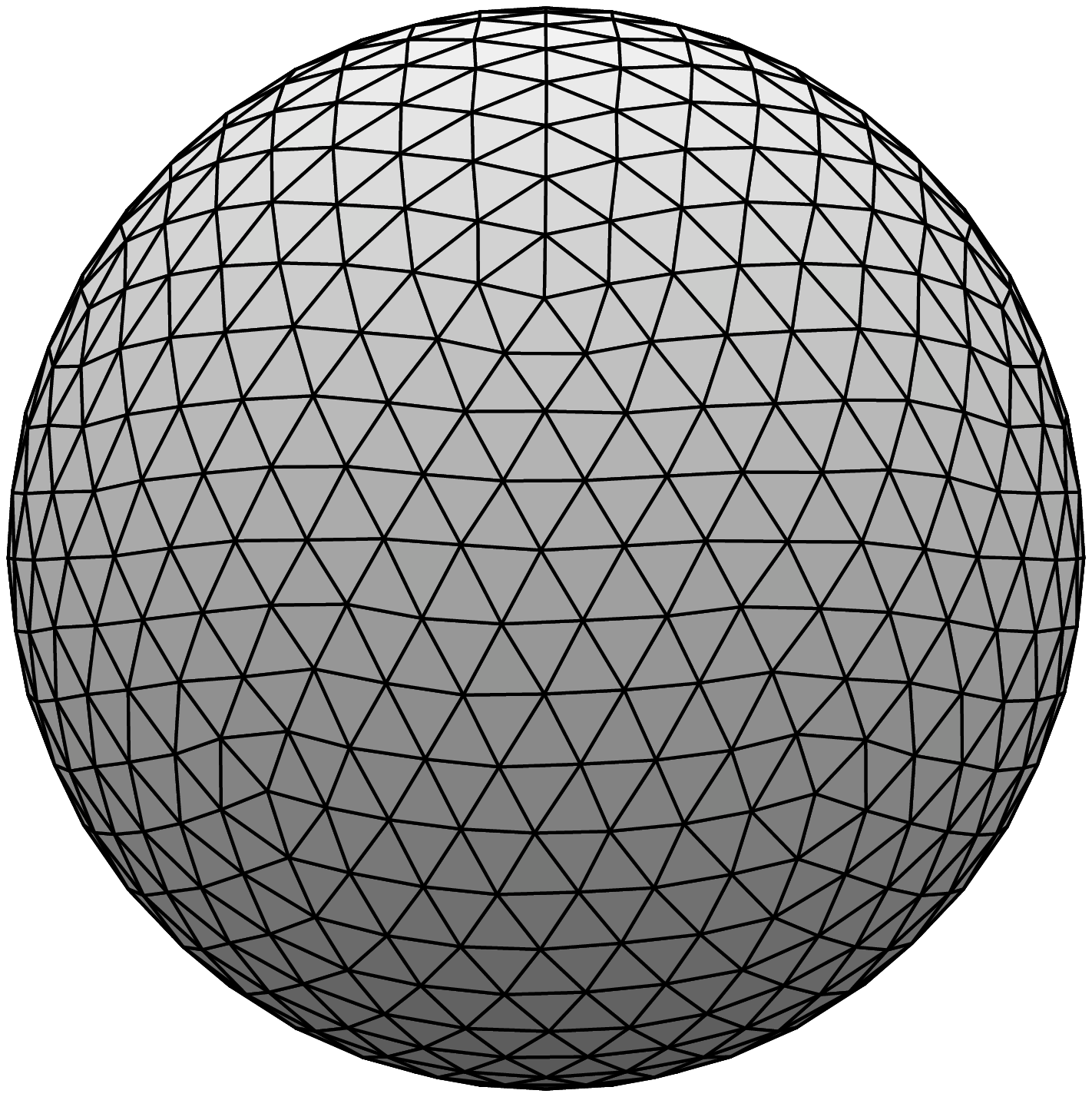} & \cc \includegraphics[scale=0.16]{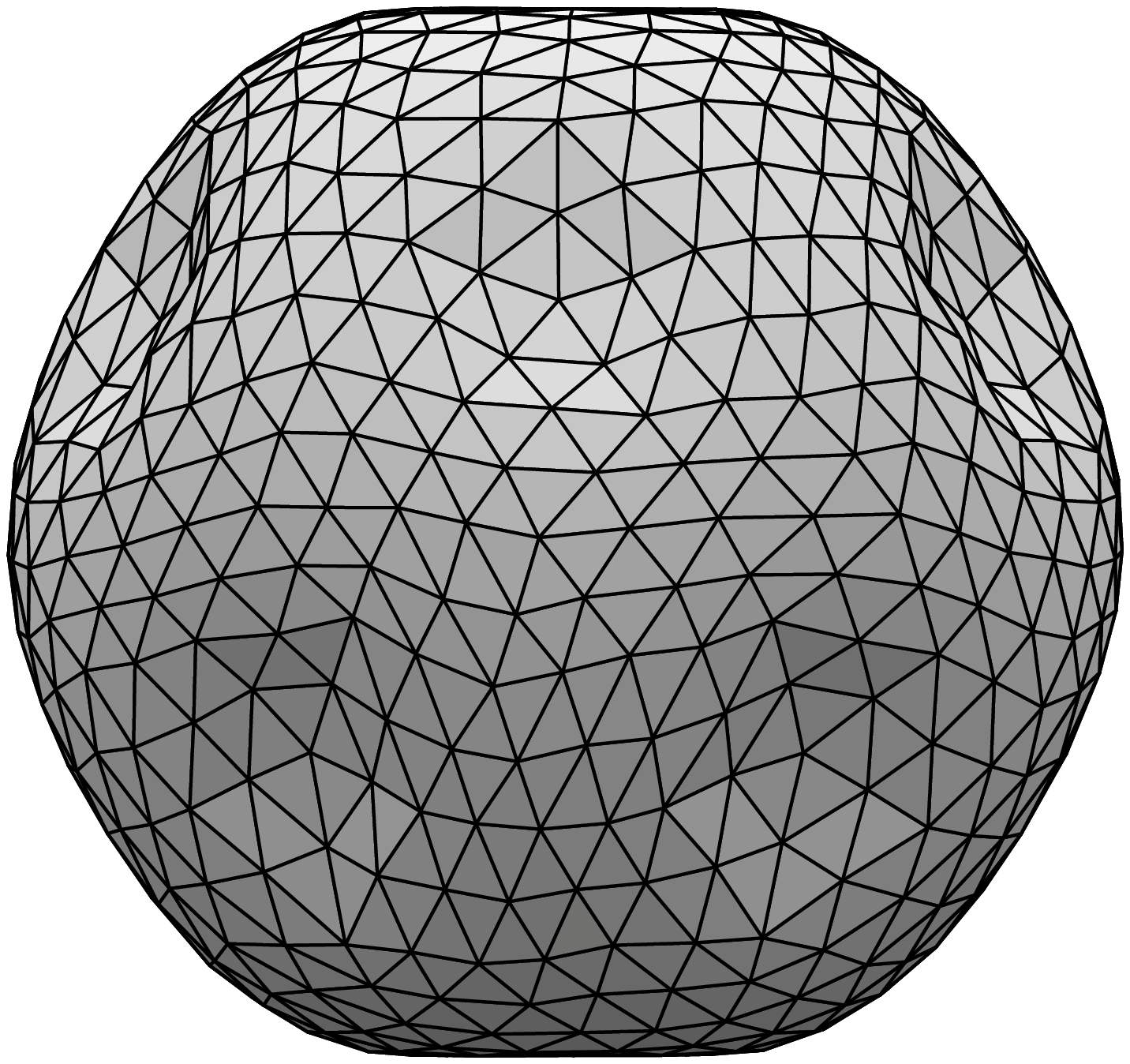} & \cc \includegraphics[scale=0.16]{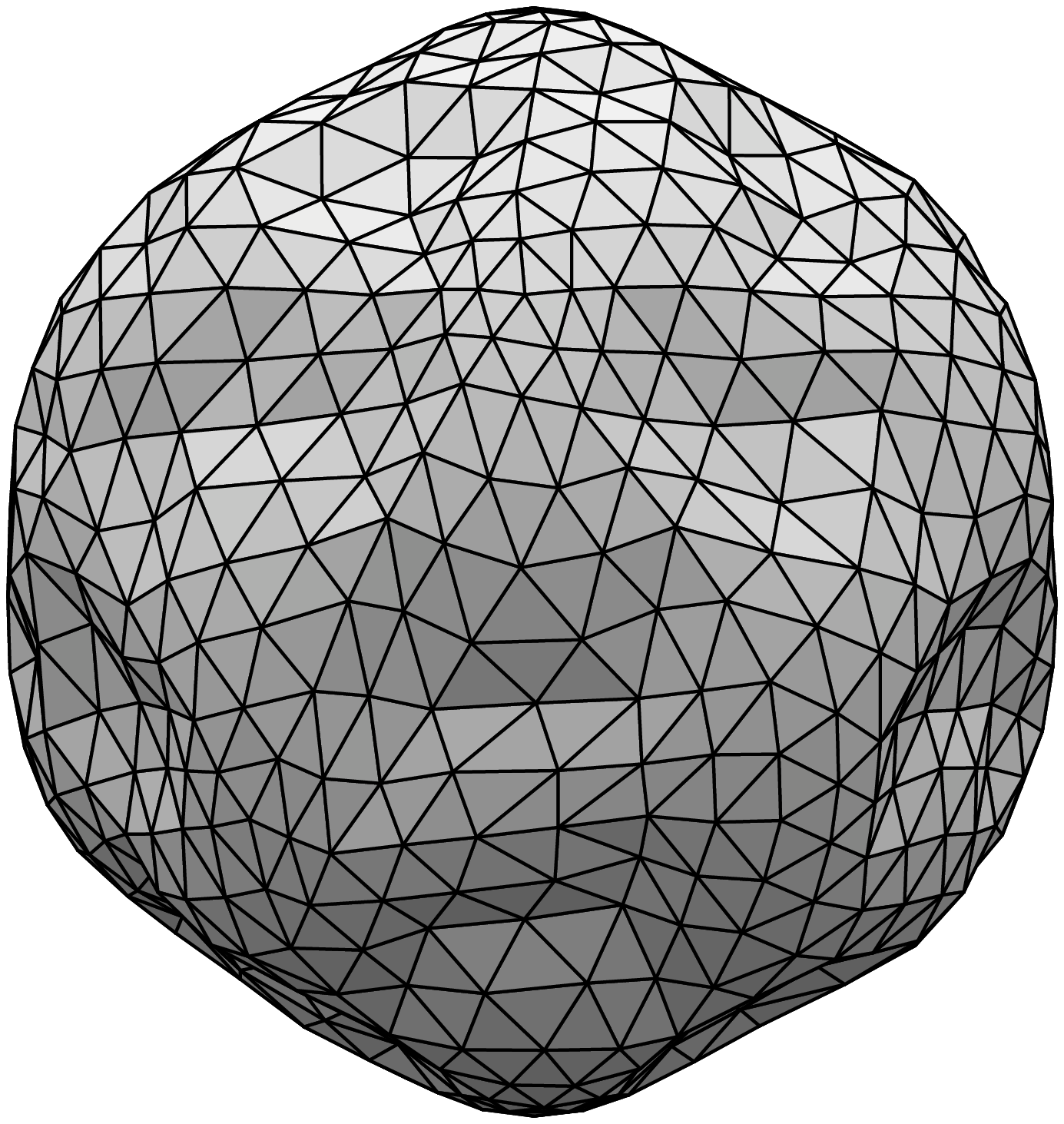}  & \includegraphics[scale=0.16]{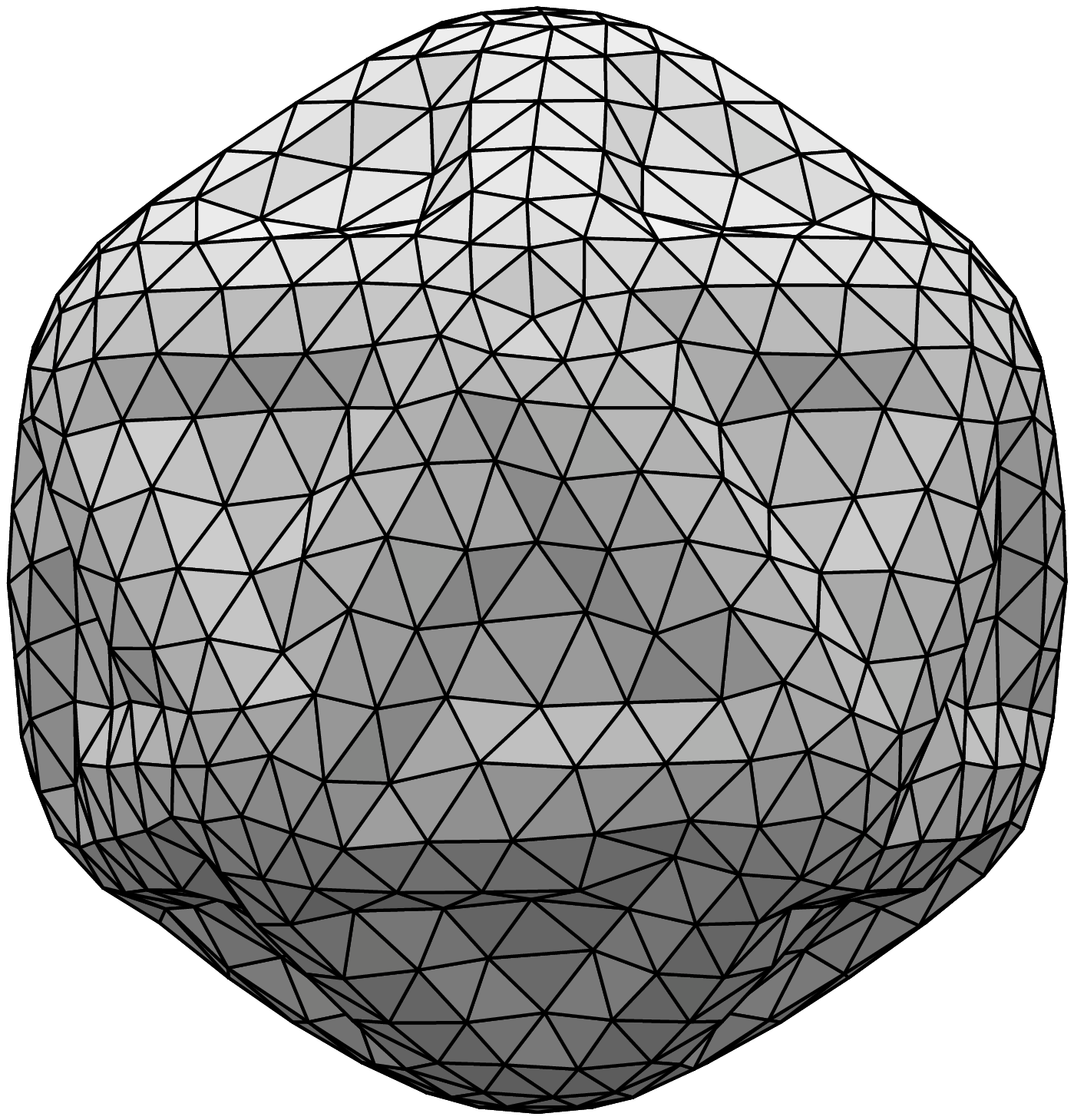} & \cc \includegraphics[scale=0.16]{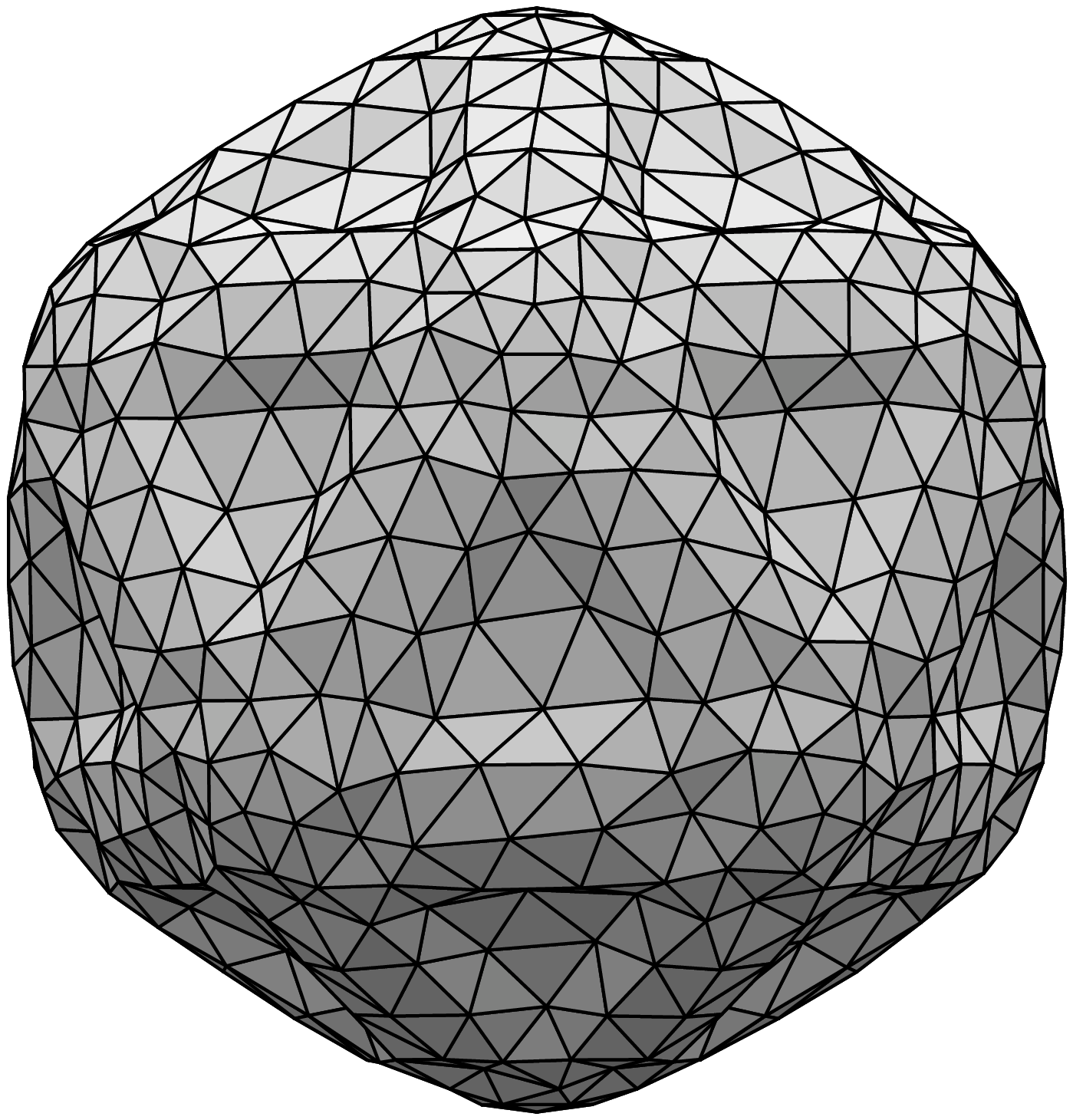} & \cc \includegraphics[scale=0.16]{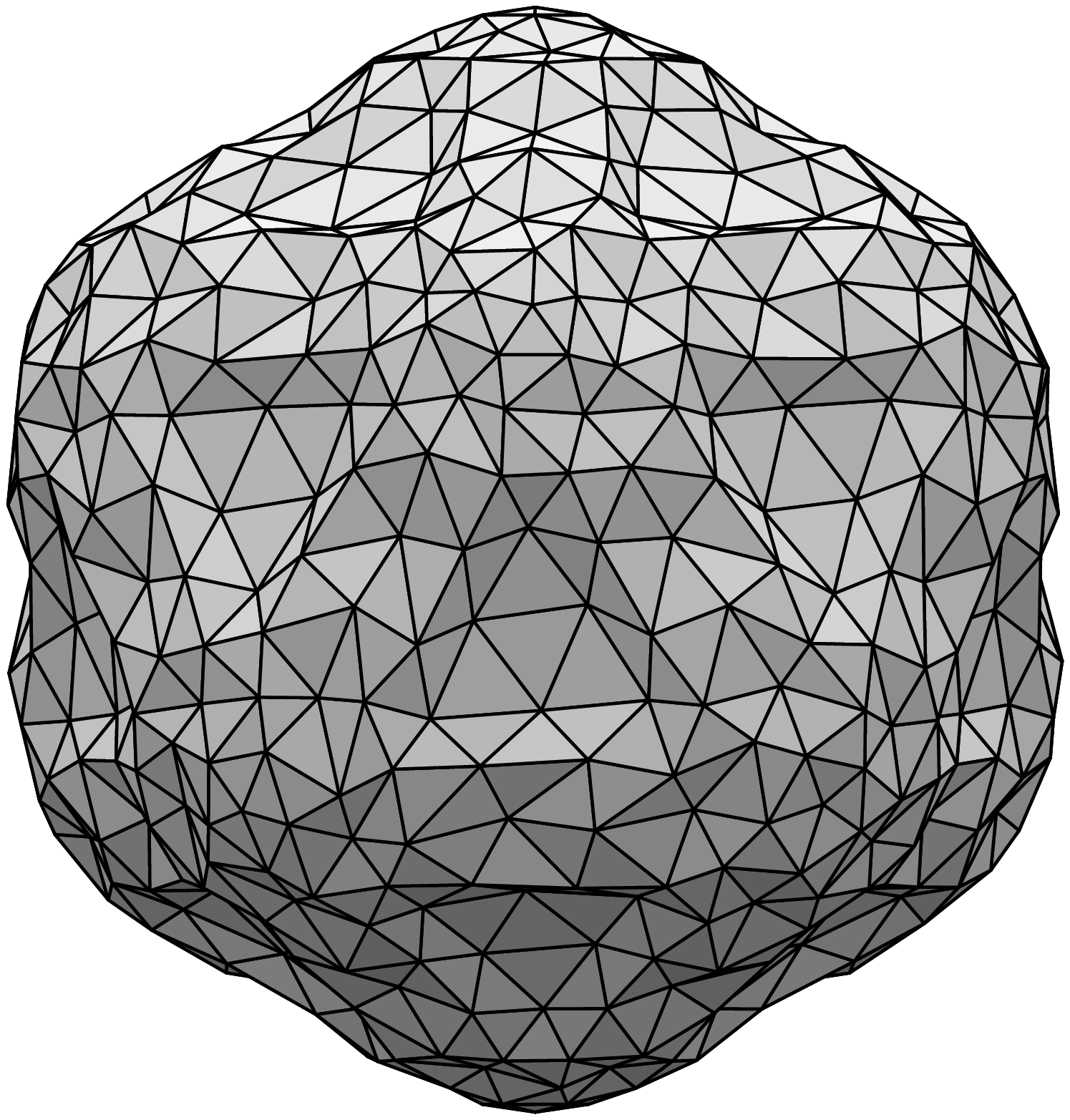}  & \\[2ex]
\cc $c_2=0$\\$c_3=0.5$ & \includegraphics[scale=0.16]{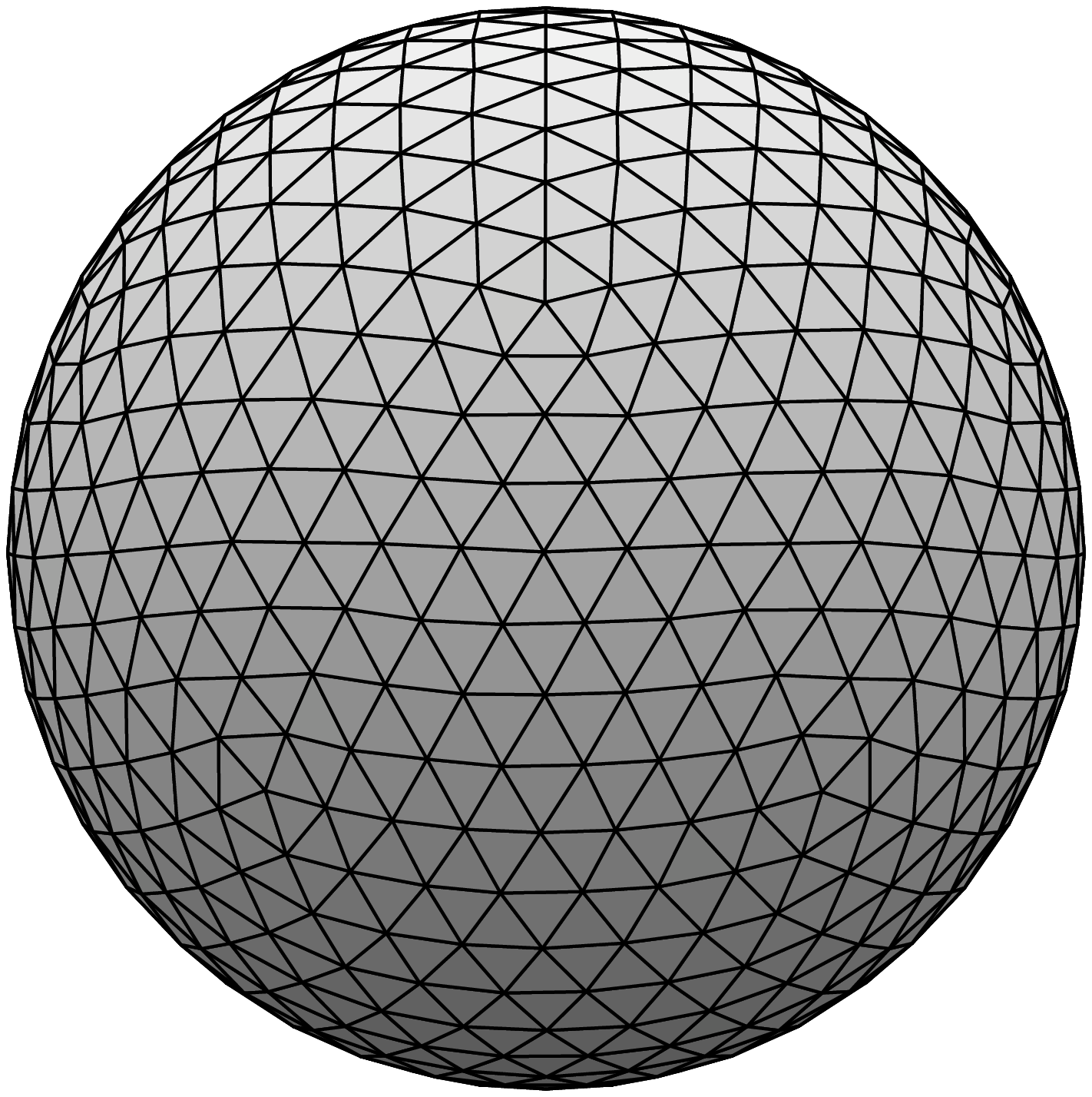} & \cc \includegraphics[scale=0.16]{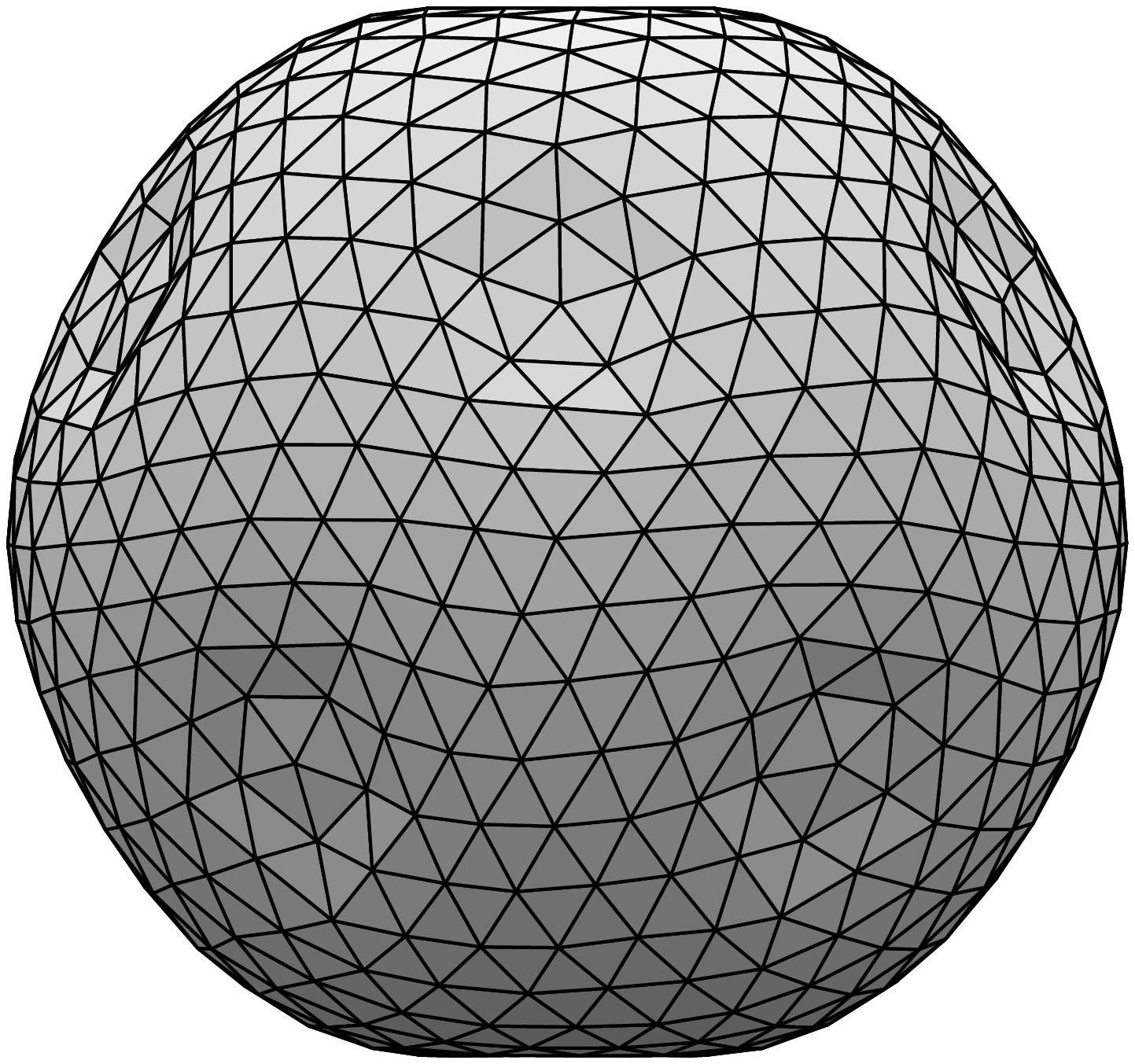} & \cc \includegraphics[scale=0.16]{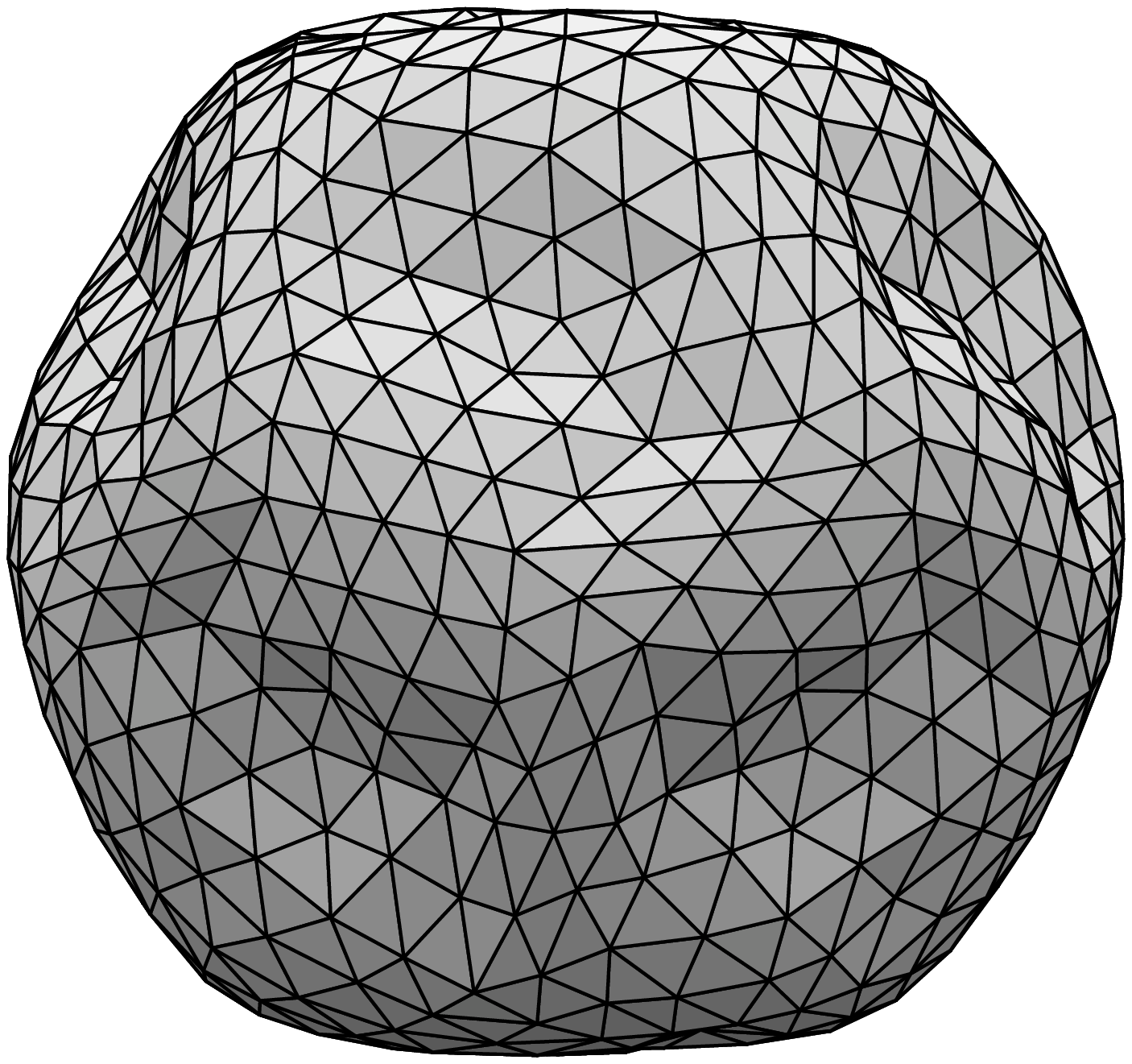}  & \includegraphics[scale=0.16]{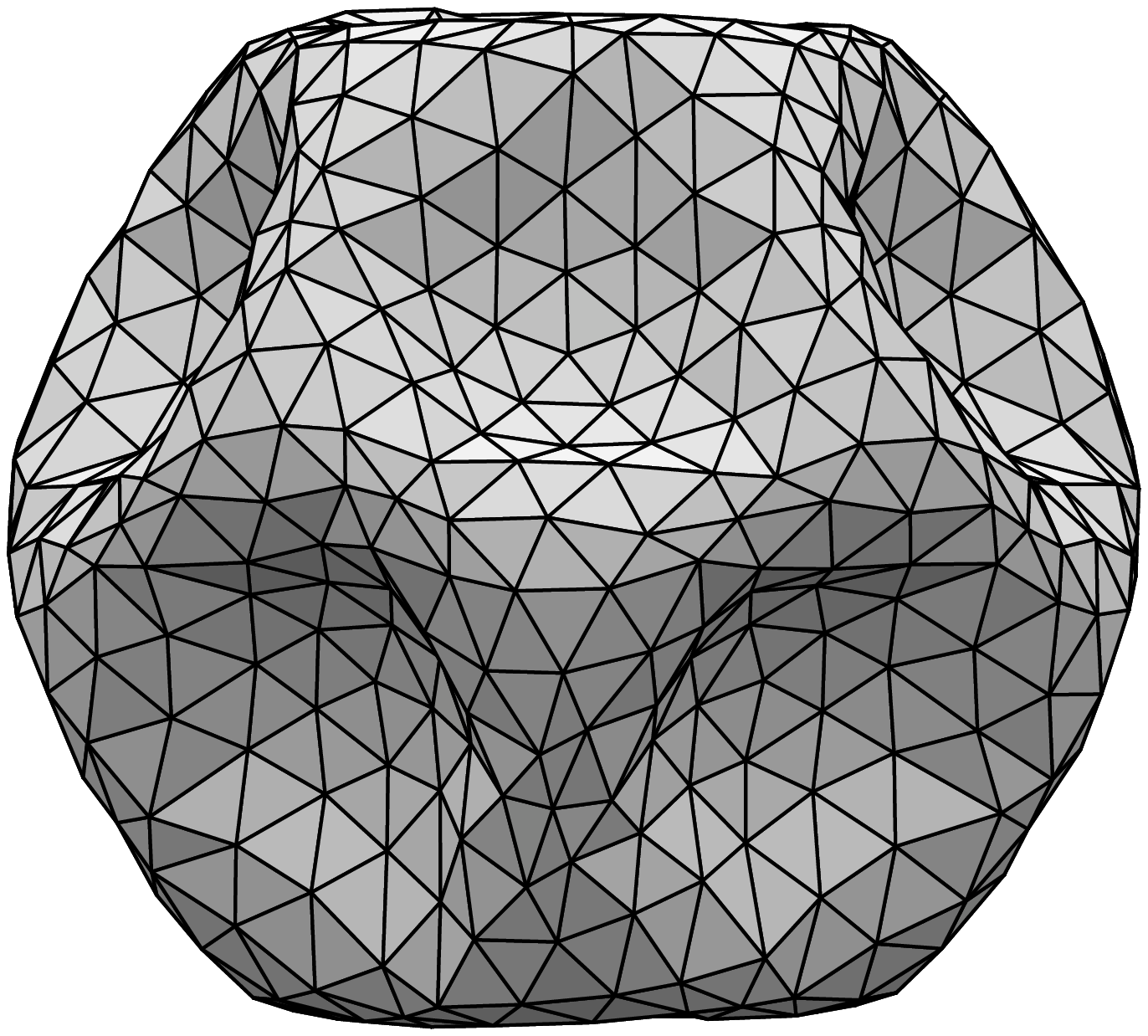} & \cc \includegraphics[scale=0.16]{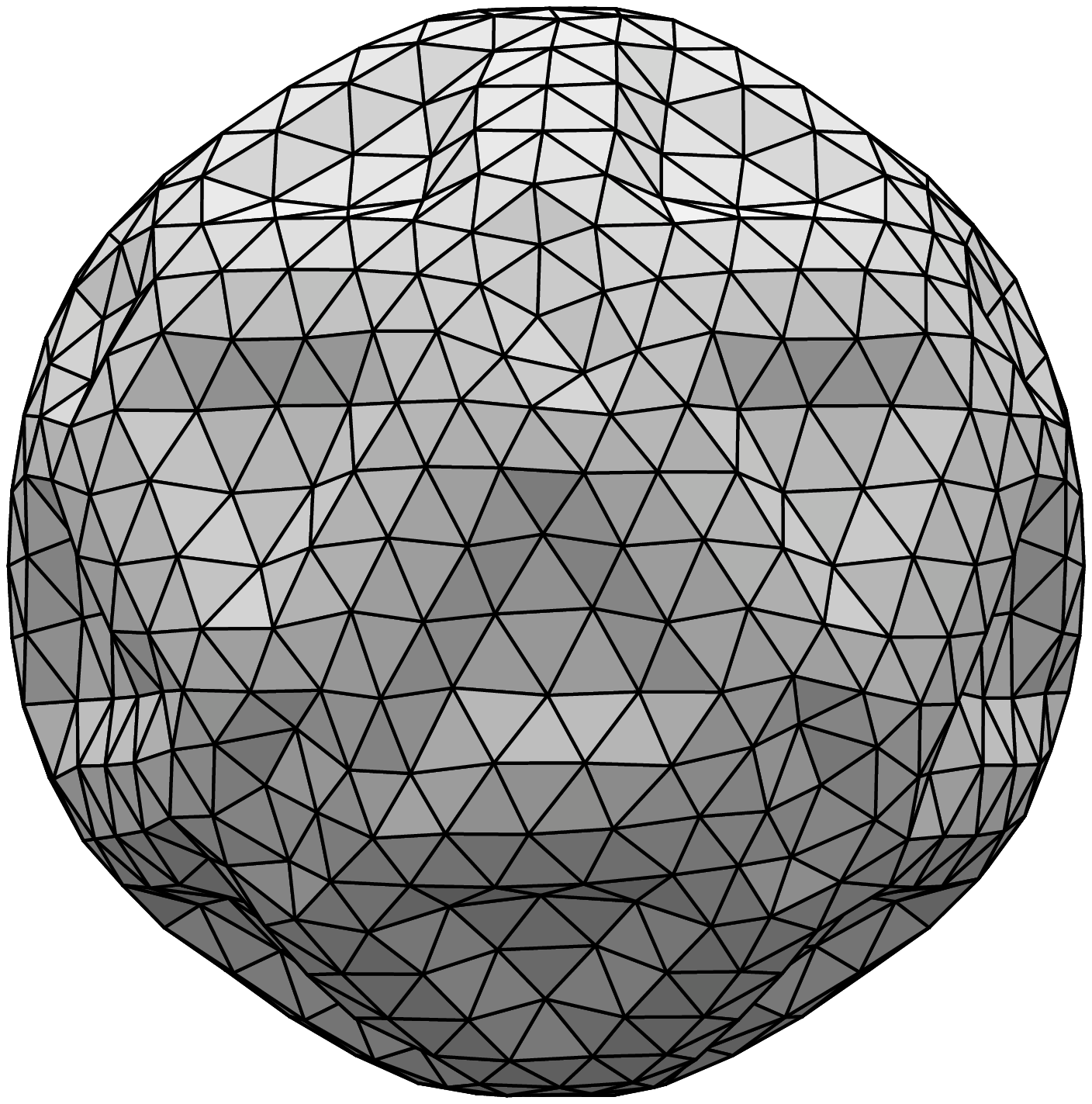} & \cc \includegraphics[scale=0.16]{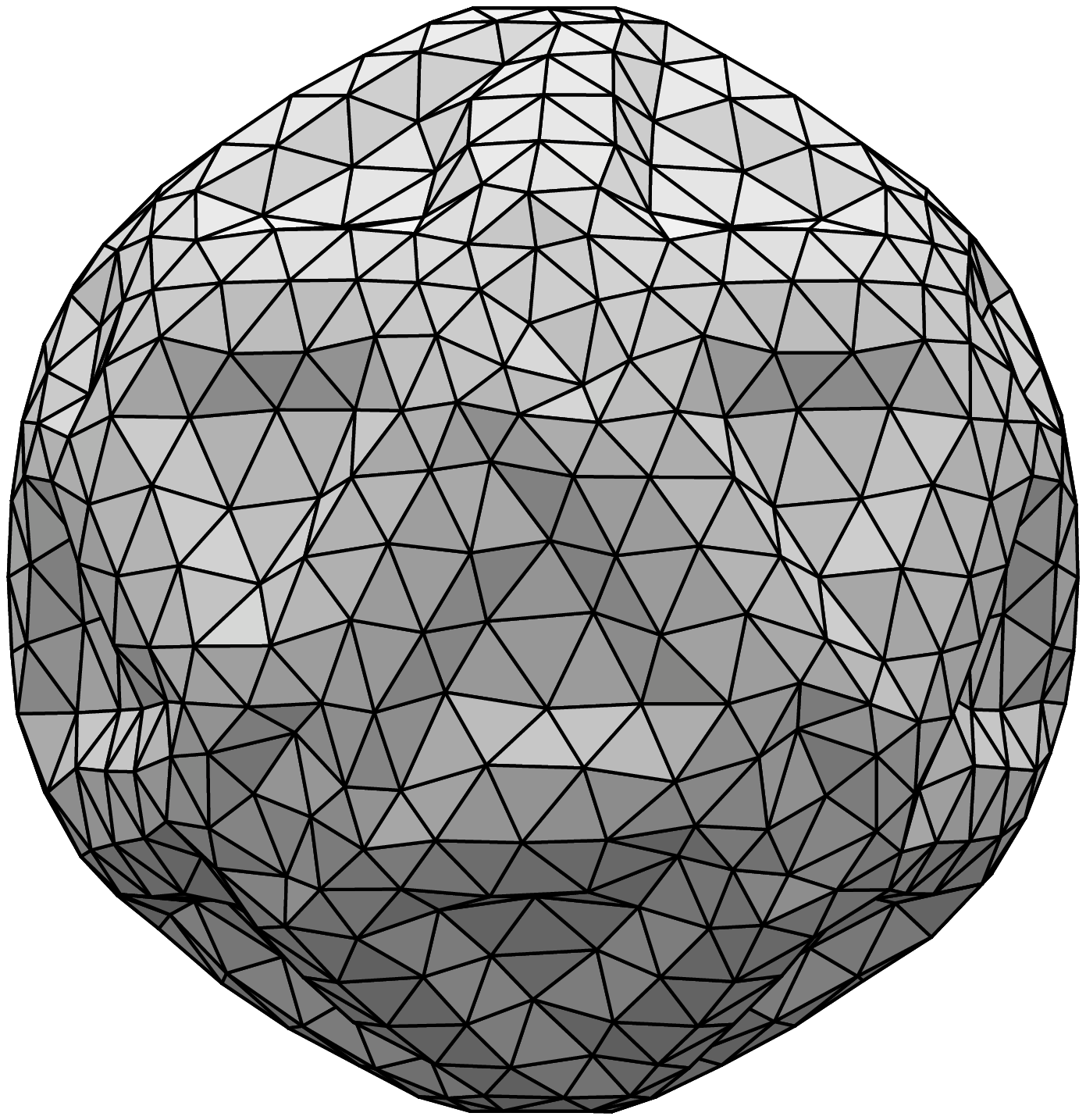}  & \\[2ex]
\cc $c_2=-1$\\$c_3=0.5$ & \includegraphics[scale=0.16]{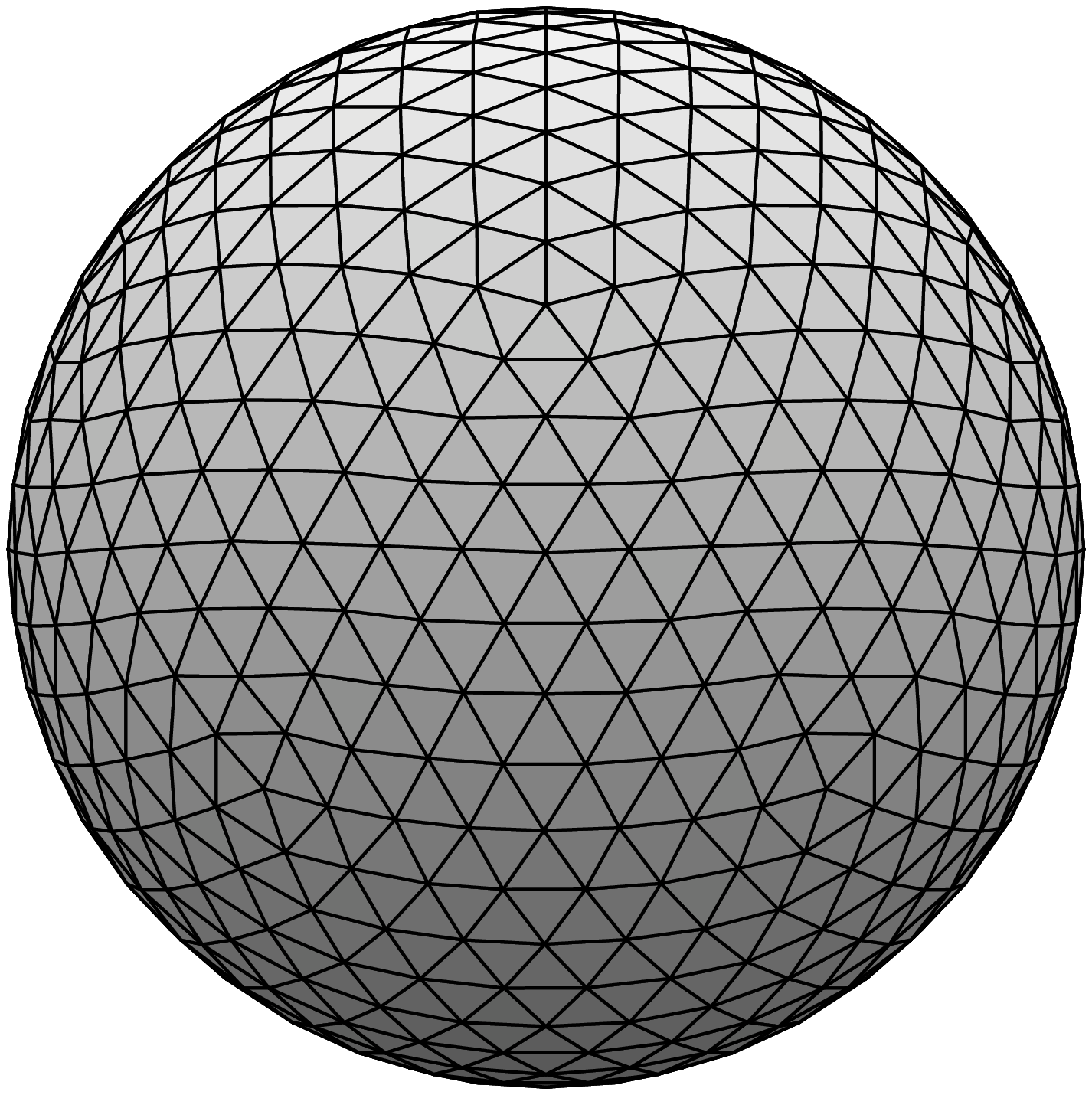} & \cc \includegraphics[scale=0.16]{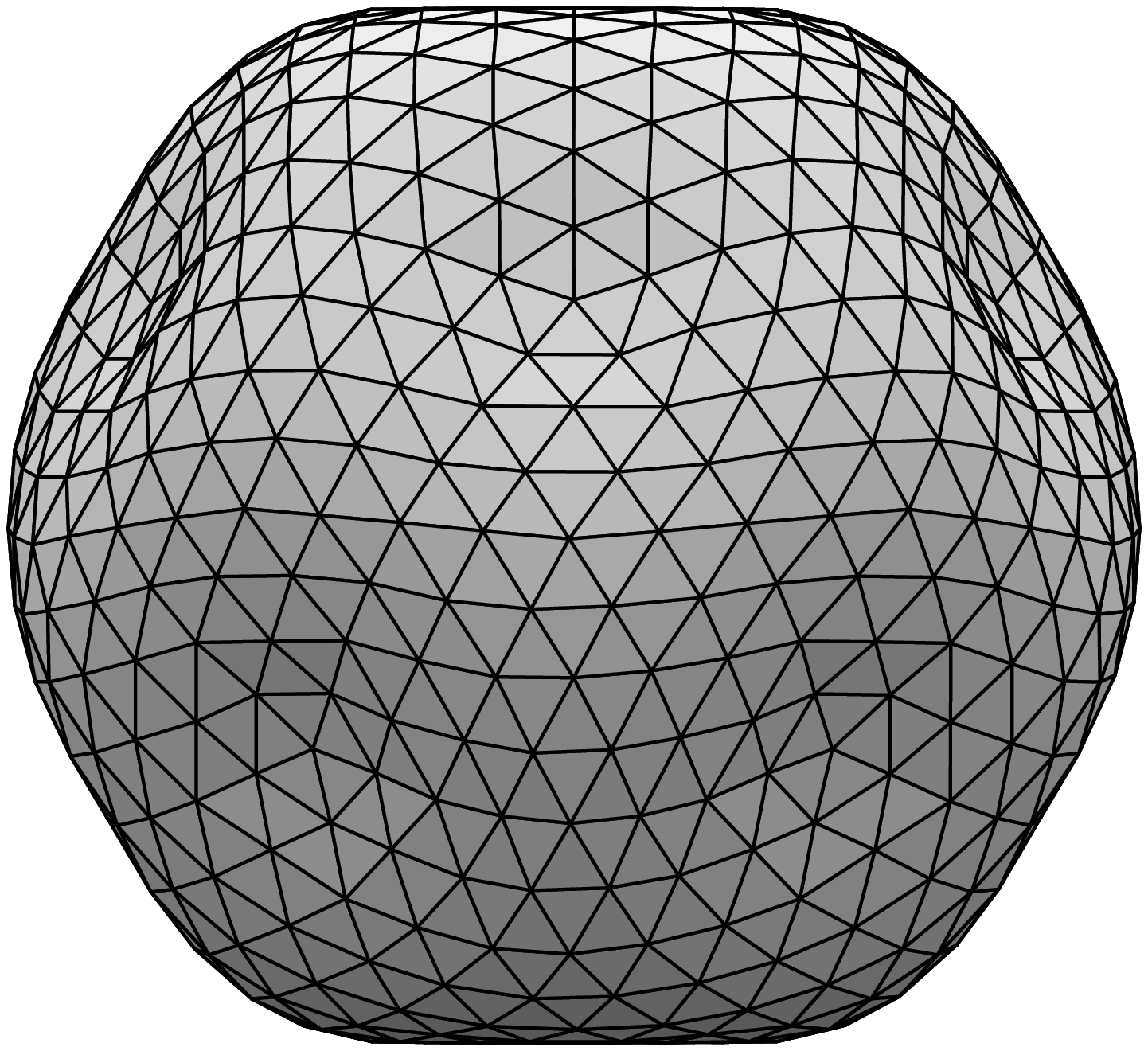} & \cc \includegraphics[scale=0.16]{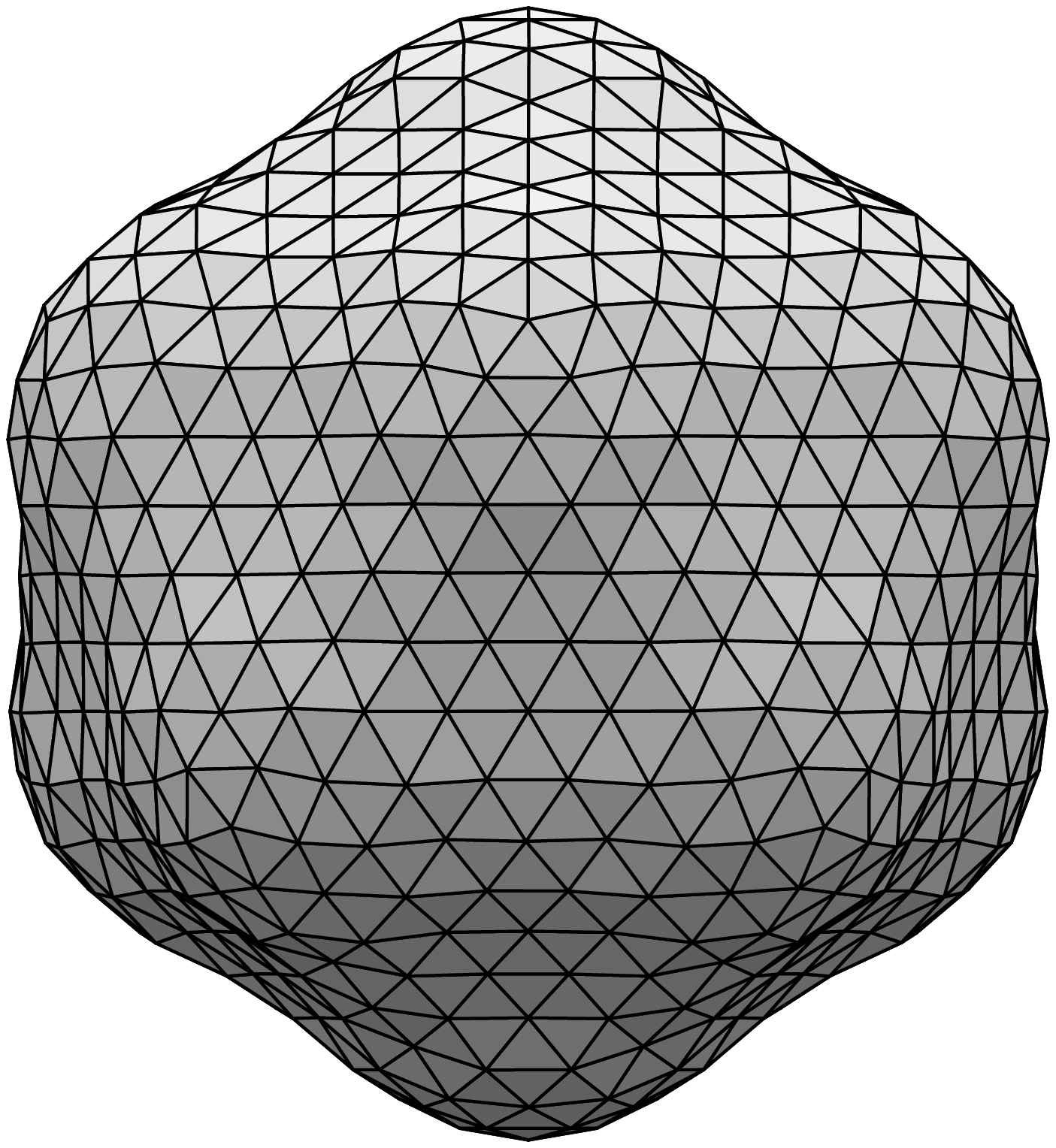}  & \includegraphics[scale=0.16]{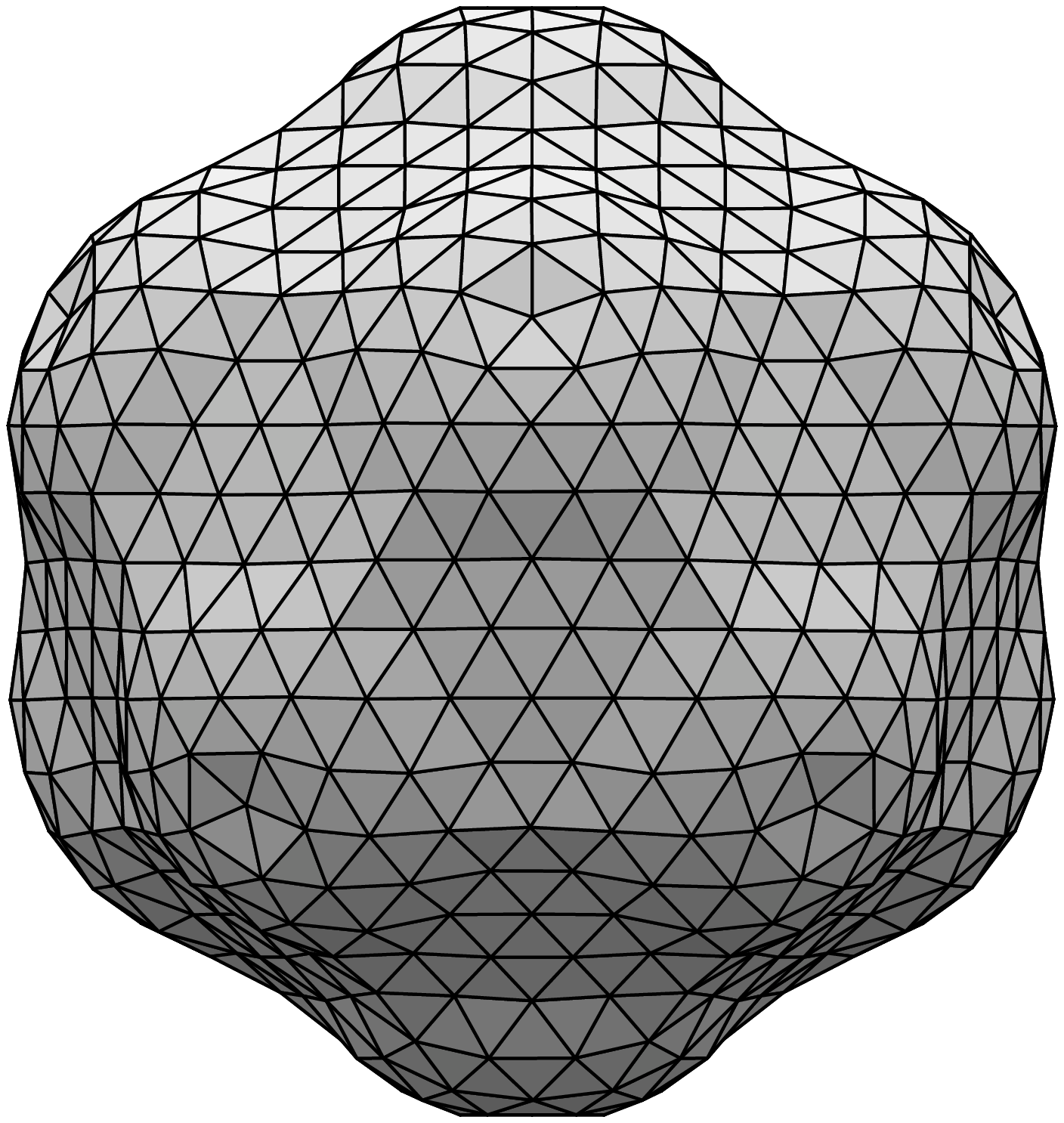} & \cc \includegraphics[scale=0.16]{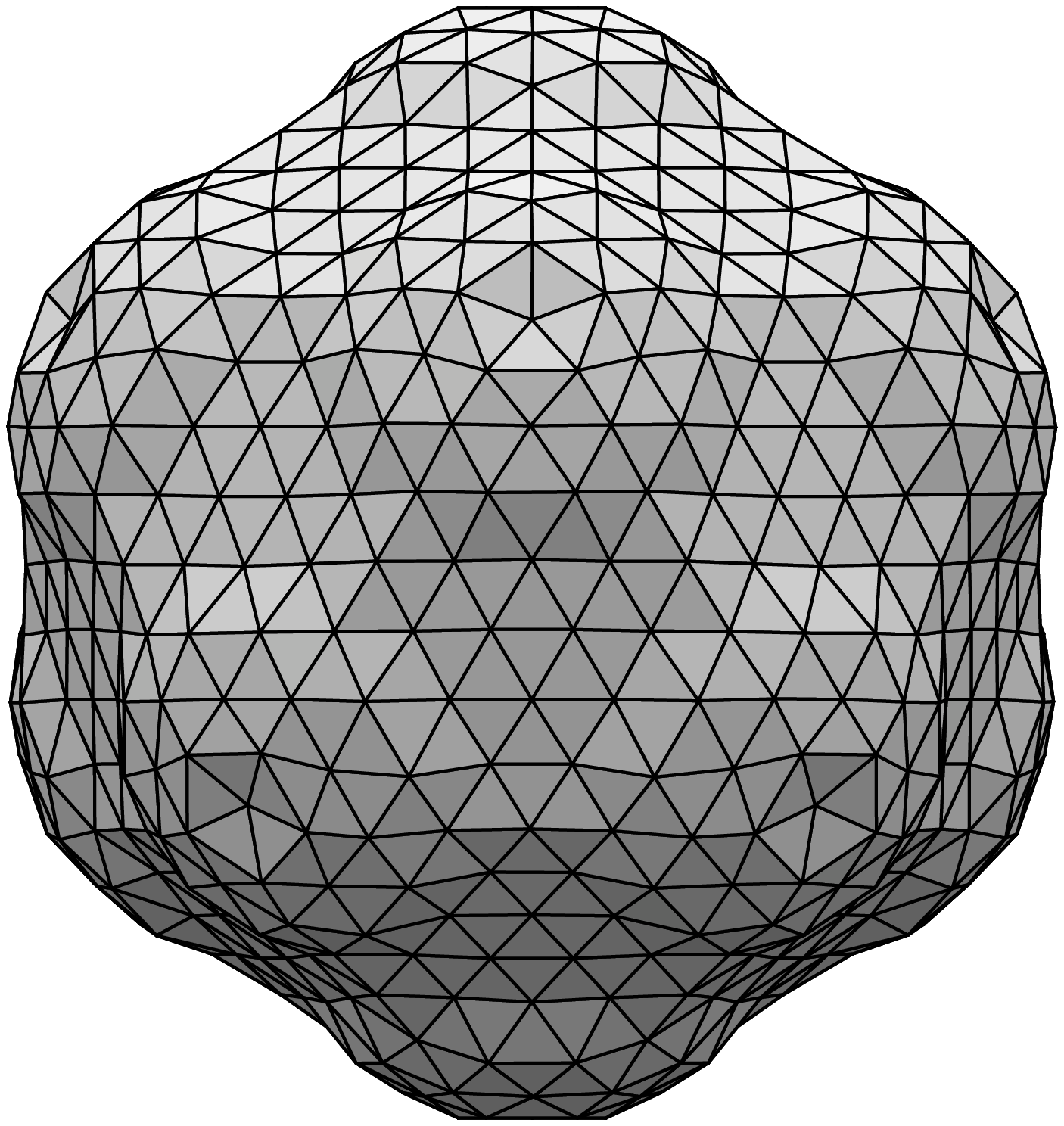} & \cc \includegraphics[scale=0.16]{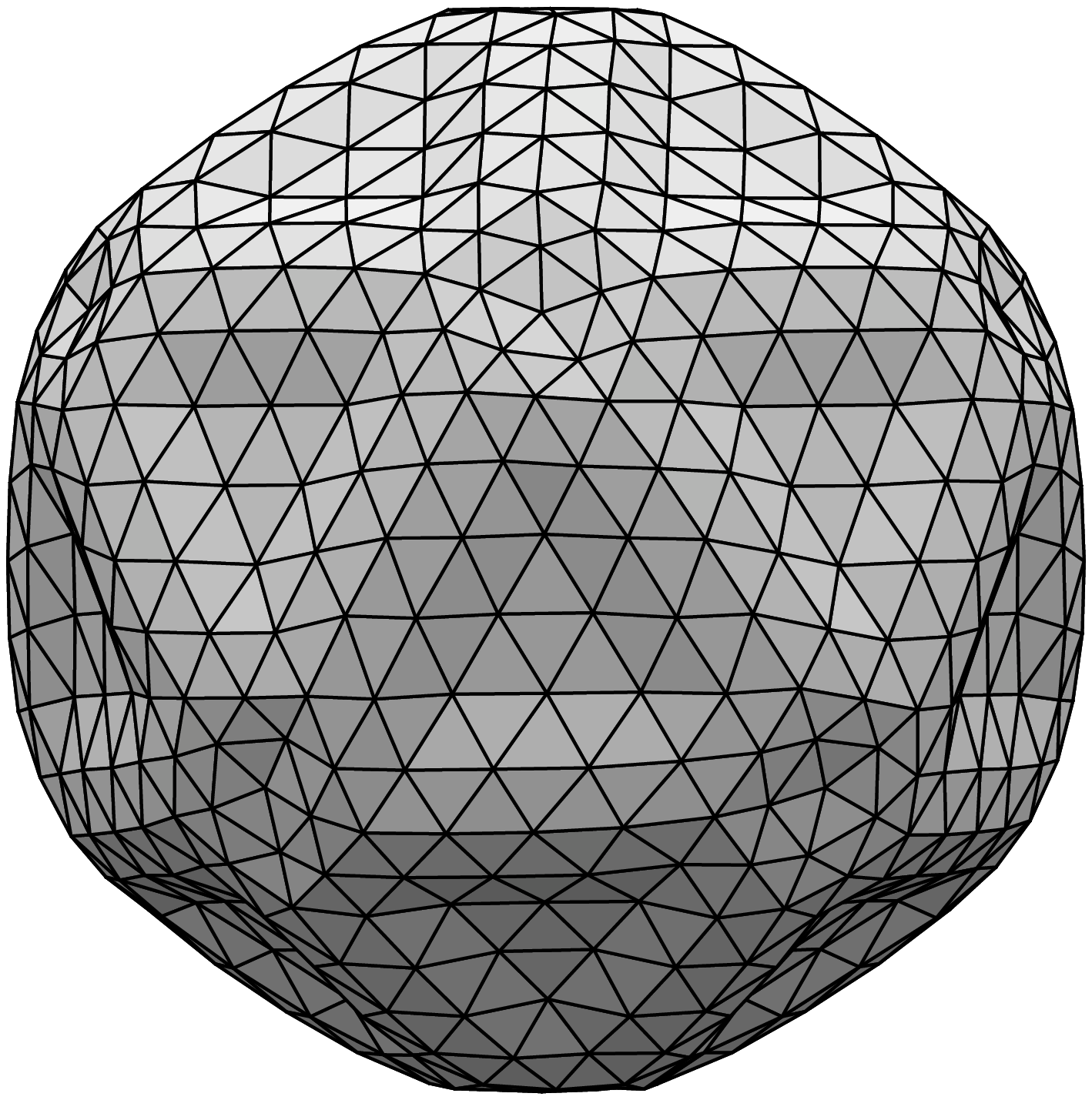}  & \\[2ex]
\end{tabular}}%
\end{table*}

Assuming a constant area of a surface $S=4\pi$  ($R=1$), we minimize the free energy given by Eq.~(\ref{eq:F4}) with``Surface Evolver''. In Table~\ref{tab1} we illustrate the equilibrium shapes   as a function of $A$ for three typical cases: 
i) $c_1=1$, $c_2=c_3=0$, ii) $c_1=1$, $c_2=0$, and $c_3=0.5$, iii) $c_1=1$, $c_2=-1$, $c_3=0.5$. Irrespective of the particular choice of the constants $c_i$, spherical shapes (shapes very close to a sphere~\cite{sphere}) are found for negative values of $A$ ($-A\iint dS\,H^2$ is positive). For $A\simeq 4$  we see a transition towards dimpled shapes (second column in Table~\ref{tab1}) followed by a transition towards shapes with ridges (last column in Table~\ref{tab1}) connecting the vertices of icosahedron. In presence of a non-zero coupling term  $c_2$ between the mean curvature $H$ and the Gaussian curvature $K$,  intermediate  shapes with bumps (see bottom row in  Table~\ref{tab1}) occur, favouring a positive $K$ for $c_2<0$. Note that, the bending rigidity $\kappa$ increases with $A$ according to Eq.~(\ref{eq:kappa}). Thus the stiffness of shapes increases in the rows, for the first and the second rows  $\kappa=A$ ($c_2=0$), for the third row $\kappa=2A$.

A canonical way to characterize  the shapes presented in Table~\ref{tab1} is to expand their radial density $R(\theta,\phi)$ in terms of spherical harmonics  $Y_{lm}(\theta,\phi)$, as follows, 
\begin{align}
R(\theta,\phi)&=\sum_{l=0}^\infty\sum_{m=-l}^l Q_{lm}Y_{lm}, \label {eq:R} 
\intertext{with the coefficients $Q_{lm}$ of the above expansion defined as} 
Q_{lm}&=\int_0^{2\pi} d\phi\int_0^\pi d\theta \sin\theta \,R(\theta,\phi)\,{Y}_{lm}^*(\theta,\phi).\label{eq:Qlm}
\end{align}
In the case of triangulated surfaces we deal with a discrete radial density $R(\theta,\phi)$, defined at the vertices of triangles.  The coefficients $Q_{lm}$ were computed  in {\it Matlab} using Gauss quadratures. Then, according to~\cite{steinhardt:1983}, we construct second and third order rotational invariants, as
\begin{align}
Q_l&=\sqrt{\frac{4\pi}{(2l+1)}\sum_{m=-l}^l |Q_{lm}|^2 },\label{eq:Ql}
\intertext{and}
W_l&=\frac{\sum_{\substack{m_1,m_2,m_3\\m_1+m_2+m_3=0}}\begin{pmatrix} l & l & l\\ m_1 & m_2 & m_3 \end{pmatrix}Q_{lm_1}Q_{lm_2} Q_{lm_3}}{\Big(\sum_{m=-l}^l |Q_{lm}|^2\Big)^{3/2}},\label{eq:Wl}
\end{align} 
where $\begin{pmatrix} l & l & l\\ m_1 & m_2 & m_3 \end{pmatrix}$ are the Wigner 3j symbols. For a sphere the only non-zero coefficient is $Q_{00} = \sqrt{4\pi}$, giving $Q_0=4\pi$.  The shapes with icosahedral symmetry are distinguished by non-zero invariants $Q_l$ and $W_l$ for $l=0,6,10,12,\ldots$~\cite{steinhardt:1983}. The vanishing of invariants for  other values of $l$ was also recovered in the present calculations. For the icosahedron, only for very high refinements, namely for triangulation with number of vertices $V=10242$, we found for $W_6$ and $W_{10}$ (notice $W_0\equiv1$) the same values as in Ref.~\cite{steinhardt:1983}. Therefore, in the following, all the integrals in Eq.~(\ref{eq:Qlm}) are computed with $V=10242$, rather than the  one presented in Table~\ref{tab1}.

\begin{figure}[t]
\includegraphics[width=0.98\linewidth]{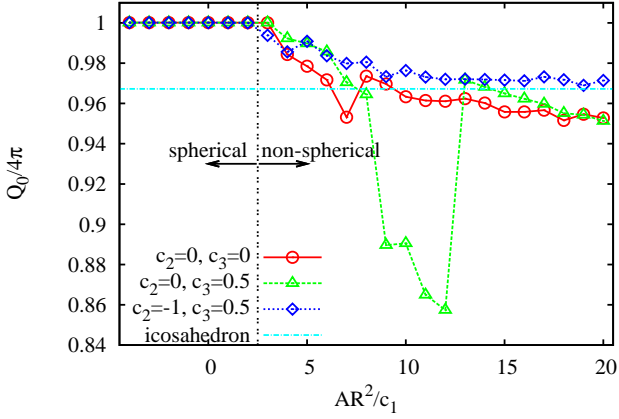}
\caption{Aspherity of the shape, described by the lowest spherical harmonic $Q_0$, normalized to the value $Q_0=4\pi$ of a perfect sphere, as a function of $A R^2/c_1$  with $c_1=1$, $R=1$ (see Eq.~\ref{eq:F4}). The vertical line indicates the crossover between spherical and non-spherical shapes.  The value of $Q_0/4\pi\simeq0.967$ for the icosahedron is shown as horizontal line.}
\label{fig:Q0}
\end{figure}

We plot first  $Q_0/4\pi$, characterizing the aspherity of the shape, as a function of the adimensional parameter $AR^2/c_1$ in Fig.~\ref{fig:Q0}. We define spherical shapes~\cite{sphere} when $Q_0/4\pi \simeq 1$, implying that  $AR^2/c_1\leqslant 2$.  Non-spherical shapes correspond to $AR^2/c_1>2$ for all three curves with different values of $c_2$ and $c_3$. The biggest deviation from a sphere,  $Q_0/4\pi\approx0.86$, happens at $A=-12$, $c_2=0$ and $c_3=0.5$. The corresponding  dimpled  shape can be found in Table~\ref{tab1}.  Notice, that the volume of the equilibrium shapes decreases with increasing $A$  in the same way as~$Q_0$. 

\begin{figure}[t]
\includegraphics[width=0.98\linewidth]{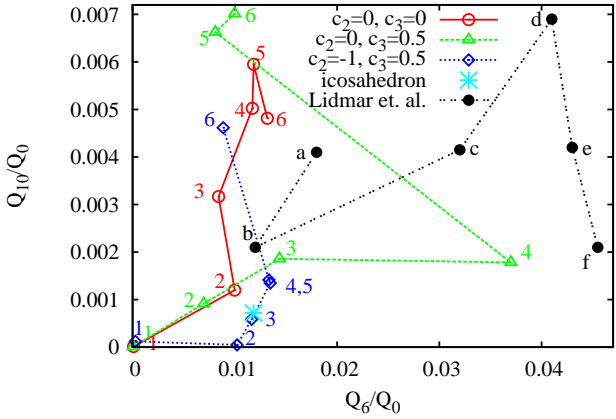}
\caption{The second order invariants defined in Eq.~(\ref{eq:Ql}). The points 1--6 correspond to the values of $A=0,4,8,12,16,20$ respectively, as in Table~\ref{tab1}. For comparison, we show the data from~\cite{lidmar:2003}, used to describe the buckling (points b--d) and faceting (points d--f) transitions of viral capsids.}
\label{fig:Q6Q10}
\end{figure}

\begin{figure}[t]
\includegraphics[width=0.98\linewidth]{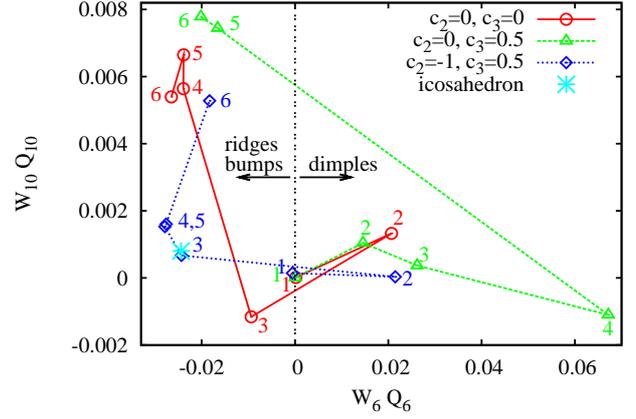}
\caption{The third order invariants $W_6$ and $W_{10}$ distinguish the shapes with bumps and ridges ($W_6<0$) from the dimpled shapes ($W_6>0$). The points 1--6 are labeled as for  Fig.~\ref{fig:Q6Q10} and  Table~\ref{tab1}. The spherical shapes  correspond to point 1 at $(0,0)$. The icosahedron is marked by an asterix.} 
\label{fig:W6W10}
\end{figure}

Figure~\ref{fig:Q6Q10} shows the second order invariants $Q_l/Q_0$ for $l=6,10$ plotted against each other. Starting from a spherical shape labeled by point (1) we follow the curves with the  points separated by $\delta A=4$, as in  Table~\ref{tab1}. The dimpled shapes correspond to the growth of $Q_6/Q_0$ (point 2 for all curves), whereas the appearance of ridges is characterized  by the increase of $Q_{10}/Q_0$ along the curves. By adding the points from Fig.~8 in Ref.~\cite{lidmar:2003}, we compare our equilibrium shapes for fluid membranes,  with shapes of crystalline viral capsids studied within the continuum elastic theory. The governing parameter of that model, relating stretching and bending of elastic shells, is the dimensionless F\"oppl-von K\'arm\'an (FvK) number $\gamma=Y R^2/\kappa$,  where $Y$ is the Young modulus. The points b--d describe the buckling transition of viral capsids, and the points d--f are associated with sharpening of the ridges at large $\gamma$~\cite{lidmar:2003}. In our case, the transition towards dimpled shapes may be analogous to the `buckling', whereas the appearance of ridges leads to the increase of $Q_{10}/Q_0$ contrary to the model of viral capsids.

To find out more connections between the non-spherical equilibrium shapes and to distinguish shapes with bumps and dimples, we consider the combination involving the third order invariants $W_l$, as shown in Fig.~\ref{fig:W6W10}. We notice that shapes with dimples are characterized by $W_6Q_6\approx0.02$, whereas shapes with bumps have the same magnitude but the opposite sign of $W_6$ (curve with rhombuses). The sign of $W_6$, which is the first non-zero third-order invariant, discriminates the shapes with dimples from the shapes with bumps and ridges. The sharpening of ridges is characterized by an increase in the invariants $Q_l$ and $W_l$  calculated for higher degree $l=10$. According to Figs.~\ref{fig:Q6Q10},~\ref{fig:W6W10} the point 3, corresponding to $A=8$, $c_1=1,c_2=-1,c_3=0.5$ (see also Table~\ref{tab1}) is the closest one to the icosahedron.


\section{Discussion and conclusions}

We have studied the equilibrium shapes of symmetric fluid membranes with a spherical topology. Assuming a fourth-order curvature model proposed in Ref.~\cite{fasolino:2006} we found a variety of shapes with dimples, bumps and ridges as well as quasi-spherical shapes. We noticed that similar shapes appear in the theory of elastic icosahedral shells, when studying the buckling and ridge-sharpening transitions~\cite{lidmar:2003,podgornik:2009}. Both these transitions depend only on the FvK number, which is the ratio of the stretching and bending contributions to the free energy. In our case, the competition arises between the negative quadratic  term in Eq.~(\ref{eq:F4}) and higher order quartic terms. Our numerical analysis shows that the transition from spherical towards dimpled shapes depends only on the value of $A$ or more likely on the dimensionless combination $AR^2/c_1$. The transitions between non-spherical shapes, such as dimples--ridges and dimples--bumps, are not determined  only by the parameter $AR^2/c_1$.  Both buckling  and ridge-sharpening transitions occur within one order of $AR^2/c_1$ whereas in the model of elastic shells the FvK number should change within four orders of magnitude~\cite{lidmar:2003,podgornik:2009}.

Our calculations were done under the constraint of  constant surface, and we found a decrease of the volume with increasing $A$, similar to the change of volume upon buckling transition~\cite{podgornik:2009}. It might be interesting to study the ${\cal F}_4$-minimizing shapes under the constraint of constant volume. With the latter constraint, non-spherical shapes appear also with the Willmore functional but those shapes are essentially different from the ones we find within our model, namely they present large deviations from spheres but no bumps, ridges or dimples.


\begin{thebibliography}{26}
\bibitem{sphere}
We avoid to use the word `sphere' on purpose, because theory predicts that a
  perfect sphere is not an equilibrium shape of ${\cal F}_4$ but we did not
  pursue the analysis of the shapes for $A<0$ in detail. However, within our
  numerical accuracy, deviations from a perfect sphere is larger for ${\cal
  F}_4$ than for the Willmore functional.
\end{thebibliography}

\end{document}